\documentstyle[12pt]{article}
\newcommand{\beq}{\begin{equation}}
\newcommand{\eeq}{\end{equation}}
\newcommand{\bea}{\begin{eqnarray}}
\newcommand{\eea}{\end{eqnarray}}
\newlength{\bredde}
\def\slash#1{\settowidth{\bredde}{$#1$}\ifmmode\,\raisebox{.15ex}{/}
\hspace*{-\bredde} #1\else$\,\raisebox{.15ex}{/}\hspace*{-\bredde} #1$\fi}
\begin{document}
%\begin{Ntitlepage}
%\docnum{CERN--TH-6959/93}
%\vspace{1cm}

\title{\Large{\bf Gauge-Symmetric Approach to Effective Lagrangians:\\
The $\eta'$ meson from QCD}}

\author{{\sc P.H. Damgaard}             \\
CERN -- Geneva \\ ~~ \\ {\sc H.B. Nielsen} \\ The Niels Bohr Institute
\\ Blegdamsvej 17, DK-2100 Copenhagen, Denmark \\ ~~ \\ and \\ ~~ \\
{\sc R. Sollacher} \\ Gesellschaft f\"ur Schwerionenforschung GSI
mbH\\ P.O.Box 110552, D-64220 Darmstadt, Germany}

\maketitle
\vfill
\begin{abstract}
We present a general scheme for extracting effective degrees of
freedom from an underlying fundamental Lagrangian, through a series of
well-defined transformations in the functional integral of the cut-off
theory. This is done by introducing collective fields in a
gauge-symmetric manner. Through appropriate gauge fixings of this
symmetry one can remove long-distance degrees of freedom from the
underlying theory, replacing them by the collective fields. Applying
this technique to QCD, we set out to extract the long-distance
dynamics in the pseudoscalar flavour singlet sector through a gauging
(and subsequent gauge fixing) of the $U(1)_A$ flavour symmetry which
is broken by the anomaly.  By this series of exact transformations of
a cut-off generating functional for QCD, we arrive at a theory
describing the long-distance physics of a pseudoscalar flavour singlet
meson coupled to the residual quark-gluon degrees of freedom. As
examples of how known low-energy physics can be reproduced in this
formulation, we rederive the Witten-Veneziano relation between the
$\eta'$ mass and the topological susceptibility, now for any value of
the number of colours $N_c$.  The resulting effective Lagrangian
contains an axial vector field, which shares the relevant features
with the Veneziano ghost. This field is responsible for removing the
$\eta'$ degree of freedom from the fundamental QCD Lagrangian.
\end{abstract}

%\vspace{35mm}

\begin{flushleft}
CERN--TH-6959/93 \\ August 1993 \\ hep-th/9308259
\end{flushleft}
%\end{Ntitlepage}
%\thispagestyle{empty}
\vfill\eject
%\pagestyle{empty}
%\clearpage\mbox{}\clearpage

%\setcounter{page}{1}
%\pagestyle{plain}
\newpage

\section{Introduction}

The concept of ``effective Lagrangians" has many different meanings,
depending on the circumstances under which they have been derived.
For example, without knowing the underlying dynamics of certain
phenomena in detail, one may try to construct a very general effective
field theory consistent with certain global symmetries such as Lorentz
invariance, CPT, flavour symmetries, etc. In doing this, it is hoped
that only a few of the terms are required to reproduce approximately
the observed phenomena, while the lack of knowledge of the underlying
theory can be absorbed into a few parameters, -- the coupling
constants of the effective Lagrangian. Fitting these couplings to
experimental observations, one has constructed a well-defined
effective theory with predictive power. Chiral Lagrangians in general
and linear sigma models are both prime examples of this approach (see,
e.g., ref.\cite{Lee} for an excellent early review). Long before the
advent of QCD, these theories accurately described a host of
strong-interaction phenomena.  Their sole basis was the inferred
spontaneously broken global $SU(N_f)_L\times SU(N_f)_R$ chiral
symmetry and related basic features of PCAC and current
algebra. Clearly, since these models were so successful in describing
strong-interaction physics in the low-energy regime, the fine details
of the underlying gauged $SU(3)$ theory of quarks and gluons were not
very important for these phenomena. A beautiful description of chiral
Lagrangians in the light of the more recent developments can be found
in a series of papers by Gasser and Leutwyler \cite{Gasser}. These
papers also contain a good introduction to the use of chiral
perturbation theory.\footnote{For more recent reviews, and
applications to anomalous processes, see $e.g.$ ref.
\cite{Ecker}.}

Today, we are of course in a different position with regards to
strong-interaction physics. We {\em know} the underlying theory -- QCD
-- at least down to distance scales that are very far removed from
hadronic phenomena. The purpose of an effective field theory for
strong interactions is then quite different. Instead of having to
solve the underlying theory exactly in the low-momentum regime
(something which in principle can be done, at least numerically), we
may wish instead to extract directly from the QCD Lagrangian those
degrees of freedom that are responsible for the long-distance
dynamics.  We do not want this in terms of very complicated non-local
objects (involving, $e.g.$, the full wave functions of multiplets of
bound states), but rather in terms of a preferably local relativistic
quantum field theory that to any given level of accuracy will describe
the low-energy phenomena. Clearly, this entails extracting from the
fundamental Lagrangian those collective degrees of freedom that
represent the spectrum of lowest-lying hadrons, and their
interactions.  In analogy with the usual notion of collective
coordinates, we shall view such hadronic excitations as {\em
collective fields} of QCD.

In general, the problem is then the following. Given a local
relativistic quantum field theory described in terms of a bare
Lagrangian valid up to a cut-off $\Lambda$, how do we conveniently
extract the relevant physics of an energy scale $M_0 \ll \Lambda$ very
far from this ultraviolet cut-off? In other words, which series of
manipulations in the path integral will leave us with a Lagrangian
that to any degree of required accuracy describes the physics at scale
$M_0$?  This can be seen as the modern viewpoint on effective
Lagrangians of more fundamental theories.  It is a viewpoint in which
one either explicitly integrates, or at least mentally {\em imagines}
integrating out all degrees of freedom between $M_0$ and $\Lambda$. In
this sense, it is a direct application of Wilson's renormalization
group ideas (see, $e.g.$, ref. \cite{Wilson}).  Weinberg's
phenomenological Lagrangians \cite{Weinberg}, and Georgi's effective
Lagrangian scheme \cite{Georgi,MaGe} are very closely related to this
point of view. For two nice expositions of the renormalization group
aspects of this approach, see also ref.
\cite{Lepage}. The price one pays for integrating out short-distance
degrees of freedom is that the full ``basis" of field operators
appears to a higher and higher extent in the low-energy
Lagrangian. This, however, is in any case to be expected of an
effective theory. But an obvious drawback of only integrating out
short-distance degrees of freedom is that it does not, at least in its
most simple formulations, lead to an effective change of variables
from, say, quarks and gluons to mesons, baryons and glueballs. The
original fermions and vector bosons get more and more ``dressed", but
their r\^{o}le as constituents of more convenient field variables does
not automatically appear.

Indeed, when it comes to the practical question of actually {\em
deriving} directly from QCD an effective long-distance theory, at
least one other approach has proven to be more successful. We are
referring here to the series of papers in which an effective chiral
Lagrangian -- describing only those low-energy degrees of freedom that
are sensitive to chiral symmetry -- has been extracted from the
transformation properties of certain generating functionals of QCD
\cite{chiral}. In a more recent paper by Espriu, de Rafael and Taron
\cite{Espriu} this idea is beautifully laid out, and carried through
to higher orders in both a derivative and $1/N_c$ expansion, $N_c$
being the number of colours. Here, for the first time, an attempt has
been made to derive a low-energy chiral Lagrangian directly from
QCD.\footnote{For related schemes, see also the contributions in
ref. \cite{Skyrmions}.} The agreement with experiment is impressive,
in particular when gluonic corrections within the $1/N_c$ expansion
are included. At no point has high-momentum modes of the QCD
Lagrangian explicitly been integrated out.  Instead, such a derivation
concerns the ``bare" Lagrangians: QCD and the corresponding meson
theory with explicit cut-offs.

With these different approaches to the derivation of effective
Lagrangians available, we should clearly not introduce yet another
method if we cannot significantly shed new light on the problem. We
believe that the technique we shall discuss in this paper stands up to
this critereon.  However, the scheme is not free of difficulties, as
we shall describe in detail below. These difficulties are however not
inherent in the -- as we call it -- collective field method itself,
but rather stem from the lack of well-defined non-perturbative
renormalization prescriptions for continuum formulations of the
underlying field theory.

The simplest way to describe our approach is to say that it provides a
systematic framework for extracting appropriate collective degrees of
freedom (our collective fields), starting directly from the underlying
Lagrangian. As we have indicated above, in a renormalization group
approach the notion of collective degrees of freedom does not
naturally appear\footnote{Although of course one may choose to
introduce such degrees of freedom by hand through the projection
operator that links the starting Lagrangian to the ``renormalized"
Lagrangian which is obtained after integrating out part of the
high-momentum modes. With such a choice, we believe one can obtain an
effective Lagrangian analogous to the one we shall derive here.}. Yet
these degrees of freedom are certainly there (and even becoming
increasingly important as one goes to larger distances), and it would
be useful to have a method with which to extract them.

A new local gauge symmetry plays a central r\^{o}le in our
method. This gauge symmetry depends on the change of variables we wish
to make, $i.e.$ implicitly on the global symmetries we choose our
collective fields to probe. It should not be confused with possible
local gauge symmetries already inherent in the starting Lagrangian
(say, $SU(N_c)$, or $SU(3)$ for QCD), nor should it be given any
physical interpretation.  It is a technical device introduced only in
order to perform certain intermediate steps. In the end, this gauge
symmetry is of course completely gauge-fixed. To demystify the meaning
of such a new local symmetry, suffice it to say that it can be
considered simply as the vehicle with which we can, in a simple and
systematic manner, perform complicated changes of variables in the
functional integral.  Some related considerations can be found in
refs.\cite{Zaks,Slavnov}.  The method has recently been tested in the
solvable realm of two-dimensional field theories where it leads to the
standard bosonization rules -- plus in fact much more
\cite{us,Ikehashi}. These exactly solvable two-dimensional examples
are very useful for showing the machinery at work in cases where we
know the final answer, and we shall return to them below.

In order to extract the collective degrees of freedom describing a
flavour singlet pseudoscalar, we are almost unavoidably led to gauging
the $U(1)_A$ symmetry.  As is well known, this symmetry is anomalous
at the quantum level, but this is of no concern here. Our chiral gauge
symmetry remains by construction {\em unbroken}, even if the original
global symmetry is broken. The collective field we shall extract from
the $U(1)_A$ transformation is related to the $\eta'$ meson in a
manner to be described below.

These general remarks should suffice as an introduction to our
approach, and after having outlined the content of the rest of this
paper, we shall proceed with the details. In section 2, we present our
effective Lagrangian scheme in a situation where the exact answer is
known (and an effective Lagrangian approach hence unnecessary), namely
the Schwinger model. The obvious analogies to QCD are made explicit by
a comparison of chiral Ward identities. Our effective Lagrangian here
shows the mass generation of the bound state of a fermion-antifermion
pair. In addition, it gives what we interpret as a Lagrangian analogue
of the ``Veneziano ghost", which plays such a crucial r\^{o}le in the
resolution of the 4-dimensional $U(1)$ problem \cite{Veneziano}.  In
section 3, we apply exactly the same procedure to 4-dimensional
massless QCD. Despite many formal analogies, the situation here is of
course far more complicated, in particular with respect to such
aspects as the r\^{o}le of the ultraviolet cut-off. However, within
the cut-off theory, we can give a reinterpretation of the
Witten-Veneziano relation \cite{Witten,Veneziano} between the $\eta'$
mass in the limit $N_c \to \infty$ and the $SU(N_c)$ pure gauge theory
topological susceptibility. We discuss the extent to which a similar
relation can be derived for any finite value of $N_c$. At intermediate
steps we show how highly non-trivial aspects of the BRST gauge-fixing
of the $U(1)_A$ symmetry conspire to leave a very simple final
result. As a byproduct, we find again a Lagrangian analogue of the
Veneziano ghost, which can be responsible for the saturation of the
relevant chiral Ward identities \cite{Veneziano}. We briefly comment
on the possibility of defining cut-off independent relations of the
same kind through an appropriate renormalization procedure. We also
discuss some of the general difficulties that are bound to arise in an
approach of this kind.

\section{A toy model: An effective Lagrangian for the Schwinger model}

To illustrate the basic ingredients of the gauge-symmetric approach to
effective Lagrangians, we shall start with a simple 2-dimensional toy
model: the Schwinger model. This model is solvable in many ways (for a
partial list of relevant references, see \cite{Schwinger}), most
notably in terms of bosonization. We shall here show how the same
information can be arrived at through the introduction of a certain
collective field $\theta(x)$ without directly bosonizing the theory.
The fact that this is possible is important, since we of course have
no hope of fully bosonizing 4-dimensional QCD.

Massless QCD and the Schwinger model have many common features. Both
theories show confinement. Usually, they are defined as an ensemble of
systems carrying integer instanton number, thus exhibiting a
non-trivial vacuum structure. At least for the Schwinger model, this
is a necessary ingredient in order to have well-defined asymptotic
states
\cite{Strocchi}; in particular, this guarantees the cluster
decomposition property for (vector) gauge-invariant objects.  The most
interesting aspect with respect to the topics we are discussing in the
following is the chiral anomaly in the $U(1)$ sector. We shall list
some of the relevant chiral Ward identities for the Schwinger model
here, and then compare in the next section with the corresponding
4-dimensional QCD analogues.

We are considering the partition function of the one-flavour Schwinger
model
\beq
{\cal Z}_{Schw} = \int\! {\cal D} [\psi ,\bar{\psi} , A_\mu]\;
e^{-\int\! d^2x\; {\cal L}_{Schw} (x)}~,
\label{eq:ZSchw}
\eeq
where we have put all (vector) gauge-fixing terms into the measure.
The Lagrangian in Euclidean space-time reads
\beq
{\cal L}_{Schw} = \frac{1}{4e^2}F_{\mu\nu}F_{\mu\nu} +
\bar{\psi}(\slash{\partial} - i\slash{A})\psi ~,
\label{eq:LSchw}
\eeq
with the (Euclidean space) conventions
\beq
\gamma_\mu = \gamma_\mu^\dagger ~~,~~ \{\gamma_\mu , \gamma_\nu\} =
2\delta_{\mu\nu}
\eeq
and
\beq
\gamma_5 = i \gamma_1 \gamma_2 = \gamma_5^\dagger ~~,~~ \{\gamma_\mu
,\gamma_5 \}=0~~.
\eeq
As mentioned at the beginning of this section, we actually have to
consider an ensemble of systems with different instanton number. In
particular this means that the measure for the vector gauge field is
given by
\beq
\int\! {\cal D}[A_\mu ] = \sum_{n=-\infty}^{+\infty} \int\! {\cal D}
[A_\mu ]_n
\label{eq:Dinst}
\eeq
where, for each $n$, the vector field $A_\mu$ is subject to the
constraint
\beq
\frac{1}{4\pi} \int\! d^2x\; \epsilon_{\mu\nu} F_{\mu\nu} = n
\label{eq:ninst}
\eeq
denoting the instanton number.

Relevant anomalous Ward identities whose derivation will be
demonstrated later are
\beq
0 = \langle \partial_\mu J_\mu^5 + \frac{1}{\pi}
\epsilon_{\mu\nu}\partial_{\mu}A_{\nu}\rangle~~,~~ J_\mu^5 =
\bar{\psi} \gamma_\mu \gamma_5 \psi ~~,
\label{eq:WI1}
\eeq
the ``anomaly equation", and
\beq
\langle \partial_\mu (J_\mu^5(x)+ K_\mu (x)) \partial_\mu (J_\mu^5(y) + K_\mu
 (y)) \rangle = \frac{1}{\pi} \partial^2 \delta (x-y)~,
\label{eq:WI2}
\eeq
the corresponding identity for the correlation function; here $K_\mu$
is defined as the topological current of the photon field,
\beq
K_\mu = \frac{1}{\pi} \epsilon_{\mu\nu} A_\nu~.
\eeq

Eqs. (\ref{eq:WI1}) and (\ref{eq:WI2}) indicate two important
properties of the Schwinger model: First of all, there is a conserved
axial current $J_\mu^5 + K_\mu$; secondly, there is an associated
Goldstone boson. The latter can be seen by introducing an
interpolating pseudoscalar field $J_\mu^5 + K_\mu = -if\partial_\mu
\varphi$ in eq.  (\ref{eq:WI2}); this implies a Green's function for
$\varphi (x)$ typical of a massless pseudoscalar field and a ``decay
constant'' $f =1/\sqrt{\pi}$. The latter being independent of the
detailed dynamics it may be misleading to call this phenomenon
``spontaneous chiral symmetry breaking''.  Another common feature of
QCD and the Schwinger model is the absence of this Goldstone boson in
the physical spectrum.  It is widely accepted that this is due to the
fact that the conserved axial current, in particular $K_\mu$, is not
gauge invariant (see, $e.g.$, \cite{Man}).

This model is solved immediately if one applies the usual
2-dimensional bosonization rules, viz.,
\beq
\bar{\psi}\slash{\partial}\psi = \frac{1}{2} \partial_\mu \sigma
\partial_\mu \sigma ~~,~~\bar{\psi}\gamma_{\mu}\psi =
\frac{1}{\sqrt{\pi}}\epsilon_{\mu\nu} \partial_\nu \sigma~,
\eeq
which allows us to rewrite the Lagrangian in the form
\beq
{\cal L}_{Schw} = \frac{1}{2}\partial_\mu
\sigma\partial_\mu\sigma + \frac{i}{\sqrt{\pi}}
\partial_{\mu}\sigma~\epsilon_{\mu\nu}A_{\nu} +\frac{1}{4e^2}
F_{\mu\nu}F_{\mu\nu}~.
\eeq
This Lagrangian can be diagonalized in a shortcut manner if one uses
the electric field strength $F_{12}$ instead of the vector field
$A_\mu$:
\beq
{\cal L}_{Schw} = \frac{1}{2} \partial_\mu \sigma
\partial_\mu \sigma - \frac{i}{\sqrt{\pi}} \sigma F_{12} +
\frac{1}{2e^2} F_{12}^2 ~.
\eeq
A simple shift
\beq
F_{12} = F_{12}' - \frac{ie^2}{\sqrt{\pi}} \sigma
\eeq
yields
\beq
{\cal L}_{Schw} = \frac{1}{2} \partial_\mu \sigma
\partial_\mu \sigma + \frac{e^2}{2\pi} \sigma^2 +
\frac{1}{2e^2} F_{12}'^2~~.
\label{eq:Ldiag}
\eeq
{}From this we read off that $\sigma$ describes a free massive
pseudoscalar of mass $m = e/\sqrt{\pi}$.  From the bosonization rules,
we note that it can be regarded as a $\bar{\psi}$-$\psi$ bound state.

Let us now consider the electric field strength $F_{12}$ in more
detail. For its Green's function we find
\bea
\langle F_{12} (x) F_{12} (y) \rangle &=& \left\langle \left( F_{12}'
(x) -
\frac{ie^2}{\sqrt{\pi}} \sigma (x)\right) \left( F_{12}' (y) -
\frac{ie^2}{\sqrt{\pi}} \sigma (y)\right) \right\rangle \cr
&=& \left\langle F_{12}' (x) F_{12}' (y) \right\rangle -
\frac{e^4}{\pi} \langle \sigma (x) \sigma (y) \rangle
\eea
In (Euclidean) momentum space this reads
\bea
G_F (p) &=& G_{F'} (p) - \frac{e^4}{\pi} G_\sigma (p) \cr &=& e^2 -
\frac{e^4}{\pi} \frac{1}{p^2 + \frac{e^2}{\pi}}
\eea
This Green's function is closely related to the topological
susceptibility,
\bea
G_{top} &=& \int d^2x \left\langle \frac{1}{2\pi}
\epsilon_{\mu\nu} F_{\mu\nu} (x) \frac{1}{2\pi}
\epsilon_{\mu\nu} F_{\mu\nu} (0) \right\rangle \cr &=&
\frac{1}{\pi^2} G_F (p=0) \cr &=& \lim_{p\to 0} \left( f^2
m^2 - \frac{f^2 m^4}{p^2 + m^2} \right) = 0~~,
\eea
with $f$ and $m$ defined above. This result is the 2-dimensional
counterpart of what Witten \cite{Witten} suggested to be the case for
QCD, namely a cancellation between a pure (vector) gauge field and a
massive pseudoscalar contribution to the topological susceptibility.

Let us now show how the same result can be obtained from our effective
Lagrangian point of view.  The first step of the construction consists
in performing a suitable field-enlarging transformation in the
functional integral (\ref{eq:ZSchw}). Wishing in the process to
introduce a flavour-singlet pseudoscalar field, call it $\theta(x)$,
an obvious choice is to perform a local chiral rotation of the quark
fields with $\theta(x)$ acting as the chiral angle:
\beq
\psi(x) = e^{i\theta (x) \gamma_5} \chi (x) \qquad , \qquad
\bar{\psi} (x) = \bar{\chi} (x) e^{i\theta (x) \gamma_5}.
\label{eq:tr5}
\eeq

Since this defines the original quark fields in terms of a local
product of new fields, it is essential that the field transformation
be made only in a regularized version of the generating functional. In
choosing the regulator we also have to be careful: It turns out that
for a chiral transformation of the kind (\ref{eq:tr5}), an
integrability condition can only be satisfied within a restricted
class of ultraviolet regulators.  This is known as the Wess-Zumino
consistency condition (see, $e.g.$, ref. \cite{Ball}), and a
regularization scheme satisfying this requirement is similarly called
consistent. A convenient consistent scheme in the fermion sector is
provided by a set of Pauli-Villars regulator fields, and this is what
we shall be using in this paper.

Having thus defined the corresponding Pauli-Villars regularized
generating functional, we can now see the consequences of the field
transformation (\ref{eq:tr5}) on this generating functional. We will
get two pieces, one ``classical" stemming from the variation of the
action in eq.(\ref{eq:ZSchw}) under the local chiral rotation, and one
``quantum mechanical" from the change of the fermionic measure. The
former part is obtained trivially. The latter, which if we reinstate
factors of Planck's constant is proportional to $\hbar$, can be
computed by means of the method described in \cite{Ball}. One central
property of this quantum mechanical part coming from the fermionic
Jacobian is that it can be rearranged into an expansion in decreasing
powers of $\Lambda$, the ultraviolet cut-off.  In this 2-dimensional
example, the cut-off can eventually be taken to infinity in a
straightforward manner, and we shall therefore only be concerned with
the leading terms in this $1/\Lambda$ expansion. (This is one of the
crucial differences between the 2-dimensional example and QCD, see
section 3 below.)

With these remarks in mind, we can write the chirally rotated
Schwinger model in the form
\beq
{\cal L}_{ext} = \frac{1}{4e^2}F_{\mu\nu}F_{\mu\nu} +
\bar{\chi}(\slash{\partial} - i\slash{A} + i\slash{\partial}
\theta\gamma_5)\chi + \frac{1}{2\pi}\partial_{\mu}\theta
\partial_{\mu}\theta - \frac{i}{2\pi}\theta~\epsilon_{\mu\nu}
F_{\mu\nu} ~.
\label{eq:Lext}
\eeq
At this stage we are able to derive the Ward identities (\ref{eq:WI1})
and (\ref{eq:WI2}) from the invariance of the partition function
(\ref{eq:ZSchw}) under variable transformations.

We have now completed the part of the field-enlarging
transformation. Next, we promote the collective field $\theta(x)$ to a
true dynamical field of the theory. In the path integral formulation
it means that we integrate over $\theta(x)$ as well. This can be done
without changing the generating functional if we note that it
corresponds to the introduction of gauge degrees of freedom, and that
these degrees of freedom need to be gauge fixed \cite{AlfDam:90}.  The
corresponding gauge symmetry reads in this case:
\bea
\chi (x) &\to& e^{i\alpha (x) \gamma_5} \chi (x) \cr
\bar{\chi} (x) &\to& \bar{\chi} (x) e^{i\alpha (x) \gamma_5} \cr
\theta (x) &\to& \theta (x) - \alpha (x).
\label{eq:2dsym5}
\eea

We have already pointed out that introducing a functional integration
over $\theta$ renders the extended partition function with the
Lagrangian (\ref{eq:Lext}) symmetric under the gauge transformation
(\ref{eq:2dsym5}).  At first sight, this may appear surprising: After
all, the fermionic functional measure is precisely {\em not} invariant
under chiral rotations such as the one of eq. (\ref{eq:2dsym5}). Does
this not spoil the gauge symmetry at the quantum level? The answer is
no: A gauge symmetry introduced in the way described here {\em
automatically} takes into account any non-trivial transformation of
the measure as well. Indeed, the {\em action} of eq.(\ref{eq:Lext}) is
not at all invariant under the chiral gauge symmetry of
eq.(\ref{eq:2dsym5}). Only if we correctly include the additional
terms from the measure is the whole expression invariant. At a more
formal level, by switching to an overcomplete basis of fields as
indicated, and integrating over all of these, one introduces a set of
Hamiltonian constraints. These constraints are the generators of the
gauge symmetry \cite{HoKi,AlfDam:90}. As we have seen, this holds in
the case of anomalous symmetries as well! (Note, incidentally, that we
are not gauging the chiral symmetry in the usual way by means of
spin-1 axial vector fields; the extra gauge degrees of freedom are
here carried entirely by the spin-0 pseudoscalar field $\theta(x)$,
the derivative of which corresponds to a ``pure gauge" in axial-vector
gauge theories.) One amusing consequence of this is that if one gauges
an anomalous global symmetry in this manner, the resulting {\em
unbroken} Ward identities derived by, $e.g.$, the residual BRST
symmetry left after gauge fixing are identical to the {\em anomalous}
Ward identities of the original global symmetry.  This holds here as
well.

In order to recover the original generating functional, we have to
gauge-fix this symmetry. We are obviously free to do this in any
manner we want. If, for example, we simply gauge-fix on the trivial
surface $\theta(x) = 0$, we just recover the starting Lagrangian. But
we are not required to make such a trivial gauge choice. Our only
constraint is that we must remove one and only one degree of freedom
-- the one we artificially added by going to a field basis enlarged by
the $\theta$ sector. We shall choose to do so by gauge fixing a
function involving fermionic bilinears.  For that purpose, it is
useful to note that the axial current shifts under (\ref{eq:tr5}) as
(for details see \cite{us})
\beq
i \bar{\psi} \gamma_\mu \gamma_5 \psi = i \bar{\chi}
\gamma_\mu \gamma_5 \chi  -
\frac{1}{\pi}\partial_\mu \theta ~~.
\label{eq:2dtrAxCur}
\eeq
The whole expression is, of course, gauge invariant; however, the
individual components on the r.h.s. are gauge dependent. The last term
in eq. (\ref{eq:2dtrAxCur}) is induced by the axial anomaly.

We use this manifestation of the chiral anomaly as a guide to choosing
a convenient non-trivial gauge fixing. Consider
\beq
\Phi =  \Delta i\pi\frac{\partial_\mu }{\partial^2}
\bar{\chi} \gamma_\mu \gamma_5\chi - (1-\Delta)  \theta
\label{eq:2dPhi}~,
\eeq
where $\Delta$ is a free parameter.  This choice of gauge implies that
the $\theta$-dependent part of the divergence of the axial current
should describe a fraction $\Delta$ of the divergence of the {\em
physical} axial current expressed through the {\em original} fermionic
degrees of freedom $\bar{\psi}$ and $\psi$.

The above gauge-fixing function is, however, defined in a very formal
manner on account of the inverse Laplacian.  Furthermore, if we wish
to implement $\Phi$ as a $\delta$-function constraint in the path
integral, we should be careful that it satisfies correct
transformation properties.  Under a global chiral rotation,
\beq
\bar{\chi}(x) \to \bar{\chi}(x)e^{i\alpha\gamma_5} ~~,~~~~~
\chi(x) \to e^{i\alpha\gamma_5}\chi(x) ~,
\eeq
the object involving $\bar{\chi}\gamma_{\mu}\gamma_5\chi$ remains {\em
classically} invariant but quantum mechanically it shifts owing to the
chiral anomaly. We find the shift similar to that for
(\ref{eq:2dtrAxCur}):
\beq
i\pi\frac{\partial_\mu}{\partial^2}
\bar{\chi} \gamma_\mu \gamma_5\chi \to
i\pi\frac{\partial_\mu}{\partial^2}
\bar{\chi} \gamma_\mu \gamma_5\chi + \alpha ~.
\eeq
The formal operator $\partial^{-2}$ is included precisely in order to
cancel the additional operator multiplying $\alpha$.

The action of the original Lagrangian ${\cal L}_{Schw}$ does not
remain invariant under constant chiral rotations, but shifts because
of the axial anomaly:
\beq
S = \int\! d^2x\; {\cal{L}}_{Schw} \to S - \frac{i}{2\pi} \alpha
\int\! d^2x\; \epsilon_{\mu\nu}F_{\mu\nu} = S - 2in\alpha~.
\eeq
We assume again that we sum only over integer winding numbers. The
action does, however, remain invariant under constant chiral rotations
of the form $\alpha = n'\pi$ with $n'$ another integer. On the other
hand, this means that $\theta(x)$ is only globally defined modulo
$\pi$.

Gauge-fixing the constant $\theta(x)$ modes is then non-trivial. The
gauge-fixing constraint must respect the above periodicity property;
there must, even in the gauge-fixed functional integral, be no
distinction between $\theta(x)$ and $\theta(x) + n'\pi$. If we choose
a $\delta$-function constraint to implement the gauge choice, this
$\delta$-function must then necessarily be {\em globally
periodic}. When fixing a definite value for $\theta(x)$, as for a
gauge-fixing function $\Phi = F(x) - \theta(x)$, this implies
\beq
\delta(F(x) - \theta(x)) = \delta(F(x) - \theta(x) - n'\pi) ~.
\eeq
This is exactly what is achieved by $\Phi$ above. Under a constant
chiral gauge transformation of magnitude $n'\pi$,
\beq
\Phi \to \Phi + \Delta n'\pi + (1-\Delta)n'\pi =
\Phi + n'\pi ~.
\eeq

Thus, if we wish to enforce the gauge choice $\Phi$ through a
$\delta$-function constraint inside the path integral, this
$\delta$-function must be globally periodic with period $n\pi$.  We
can represent such a globally periodic $\delta$-function by means of a
Nakanishi-Lautrup auxiliary field $b(x)$, so that this gauge-fixing
function just provides a few new terms in the action.

In practice, this is done by implementing the globally periodic
$\delta$-function using a functional Fourier representation:
\bea
\delta(\Phi) &=& \int\! {\cal{D}}[b]\; \exp\left[+\int\! d^2x d^2z\; b(x)
\left(i\pi\Delta\partial^{-2}_{(x-z)}
\partial_{\mu}^{(z)}j^5_{\mu}(z) - (1-\Delta ) \delta(x-z) \theta(z)\right)
\right] \cr
&=& \int\! {\cal{D}}[b]\; \exp\left[+\int\! d^2x\; \left(i\pi\Delta
B_{\mu}(x)j^5_{\mu}(x) + (1-\Delta )\theta(x) \partial_{\mu}
B_{\mu}(x)
\right)\right]
\label{eq:FTgauge}
\eea
Here, $j_\mu^5 = \bar{\chi} \gamma_\mu \gamma_5 \chi$ denotes the
axial vector current of the transformed fermions; the axial vector
field $B_{\mu}(x)$ is defined by
\beq
B_{\mu}(x) \equiv \int\! d^2y d^2z\; b(y)\partial^{-2}_{(y-z)}
\partial_{\mu}^{(z)}\delta(z-x) ~.
\eeq
Note that this implies $b(x) = -\partial_{\mu}B_{\mu}(x)$.  A globally
periodic $\delta$-function can be represented by
\beq
\delta (\Phi (x)) = \sum_{k=-\infty}^{\infty} \int\! {\cal D}
[b]_k\; \exp\left[ \int\! d^2x\; b(x) \Phi (x)\right]~,
\eeq
where $b$ is constrained:
\beq
\int\! d^2x\; b(x) = -\int\! d^2x\; \partial_{\mu}B_{\mu} = 2ik ~.
\eeq
This global constraint means that $b(x)$ (or $\partial_{\mu}B_{\mu}$)
share certain properties with topologically non-trivial fields
\cite{us}.

The gauge-fixing is complete once the appropriate ghost term is
found.  This can be achieved by requiring BRST symmetry in the
gauge-fixed Lagrangian. In contrast with the usual situation, the
functional measure is not invariant under the BRST transformations
\begin{eqnarray}
\delta \bar{\chi}(x) &~=~& i\bar{\chi}(x)\gamma_5c(x) \cr
\delta \chi(x) &~=~& -ic(x)\gamma_5\chi(x) \cr
\delta \theta(x) &~=~& -c(x) \cr
\delta c(x) &~=~& 0 \cr
\delta \bar{c}(x) &~=~& b(x) \cr
\delta b(x) &~=~& 0~~.
\end{eqnarray}
This property, however, is responsible for the gauge invariance (and
therefore BRST invariance) of the partition function of the modified
Lagrangian ${\cal{L}}_{ext}$.  The only BRST-variant term is therefore
the one that has been added to the modified ${\cal{L}}_{ext}$ by
imposing the gauge constraint. Its BRST variation is
\beq
\delta \left[ b(x) \Phi (x) \right]= b(x)c(x)  ~~.
\eeq
{}From that we immediately see that we need a trivial ghost term,
namely
\beq
{\cal{L}}_{ghost} = \bar{c}c~,
\eeq
whose BRST variation precisely cancels the one above. This is just
what one obtains when BRST gauge-fixing in the standard manner by
adding a term $\delta[\bar{c}\Phi]$ to the action.  Obviously, the
ghosts decouple and will be neglected in the following.

The complete gauge fixed Lagrangian now reads
\bea
{\cal L}'_{Schw} &=& \bar{\chi} \biggl( \slash{\partial} -i\slash{A}
-i\Bigl( \pi\Delta\slash{B} -\slash{\partial}
\theta
\Bigr) \gamma_5\biggr) \chi + \bar{c} c + \frac{1}{4e^2}F_{\mu\nu}
F_{\mu\nu} \cr && +\frac{1}{2\pi}\partial_\mu \theta
\partial_\mu \theta -
\frac{i}{2\pi}\theta\epsilon_{\mu\nu}F_{\mu\nu} - (1-\Delta) \theta
\partial_\mu B_\mu  ~.
\label{eq:2dLgf}
\eea
{}From now on, let us concentrate on the gauge $\Delta = 1$, which in
the language of ref. \cite{us} is called the ``bosonization gauge".

The bosonization of the Schwinger model is now a one-step
procedure. Shifting $B_\mu$ as
\beq
B_\mu \to B_\mu + \frac{1}{\pi} \partial_\mu \theta + \frac{i}{\pi}
\epsilon_{\mu\nu} A_\nu ~,
\eeq
thereby taking advantage of the relation $\gamma_\mu
\gamma_5 = i\epsilon_{\mu\nu} \gamma_\nu$ for 2-dimensional
$\gamma$-matrices, the fermions decouple completely from $A_\mu$ and
$\theta$. The Lagrangian now reads
\bea
{\cal L}_{bos} &=& \bar{\chi} ( \slash{\partial} -i\pi\slash{B}
\gamma_5 ) \chi + \bar{c} c \cr && + \frac{1}{4e^2}F_{\mu\nu}
F_{\mu\nu} +\frac{1}{2\pi}\partial_\mu \theta
\partial_\mu \theta -
\frac{i}{2\pi}\theta\epsilon_{\mu\nu}F_{\mu\nu}
\label{eq:Lbos}
\eea
The fields $\bar{\chi},\chi ,B_\mu, \bar{c}$ and $c$ can now be
integrated out; this provides a normalization factor necessary to
yield the correct normalization with respect to the original fermionic
representation of the partition function (\ref{eq:ZSchw}).

The careful reader may wonder whether we are allowed to perform the
shift of $B_\mu$. We recall that in general $A_\mu$ can be written as
\cite{Jaye}
\beq
A_\mu (x) = C_\mu (x) + \epsilon_{\mu\nu} \partial_\nu \phi (x) +
\partial_\mu \varphi (x)
\eeq
where $C_\mu$ carries the topological information, $i.e.$
\beq
\frac{1}{2\pi}\int\! d^2x\; \epsilon_{\mu\nu} \partial_\mu C_\nu
= n~~.
\eeq
The term $\partial_\mu \varphi (x)$ is a pure (vector) gauge
component; it can be removed either by a corresponding (vector) gauge
transformation of the fermion fields or by a suitable (vector) gauge
condition. The shift of $B_\mu$ then means
\beq
B_\mu \to B_\mu + \frac{1}{\pi} \partial_\mu \theta -
\frac{i}{\pi} \partial_\mu \phi +\frac{i}{\pi} \epsilon_{\mu\nu} C_\nu
{}~~.
\eeq
This implies
\beq
-\partial_\mu B_\mu = b \to b- \frac{1}{\pi} \partial^2 \theta +
\frac{i}{\pi} \partial^2 \phi - \frac{i}{\pi} \epsilon_{\mu\nu}
\partial_\mu C_\nu
\eeq
and therefore, after integration over euclidean space-time,
\beq
2ik \to 2ik - 2in~~.
\eeq
The latter transformation is just a shift of the summation index $k$
by the instanton number $n$ defining a new summation index $k'$. This
is a very important point: only with the gauge fixing with a globally
periodic $\delta$-function do we arrive at the bosonized version of
the Schwinger model. Any other choice of gauge would presumably {\em
not} lead to a {\em complete} decoupling of the fermions from the
vector field $A_\mu$. Another aspect of this gauge will be discussed
in a moment.

We are now presenting a more detailed investigation of the mass
generation mechanism starting again from the Lagrangian
(\ref{eq:2dLgf}). The idea is to construct an effective Lagrangian for
the field $\theta$ using as few properties special to two dimensions
as possible.  For this purpose we take again (\ref{eq:2dLgf}) with
$\Delta =1$ and integrate out all fields except $\theta$. The part of
the action containing the coupling of $\theta$ to the vector gauge
field and to the fermionic axial current, denoted by $X$ in the next
equation, is treated in a cumulant expansion,
\beq
\langle e^X \rangle = e^{\langle X\rangle +\frac{1}{2}
\langle X^2 \rangle_c + \ldots }
\label{eq:cum}
\eeq
where $\langle X^2 \rangle_c = \langle X^2 \rangle - \langle
X\rangle^2 $ denotes the ``connected'' part of the corresponding
moment. The expectation values $\langle \ldots \rangle$ have to be
taken with respect to a {\em truncated} theory where $\theta$ is set
to zero. In the following, we will indicate this by an index
$_{trunc}$. Furthermore, due to the gauge constraint represented by a
coupling between $B_\mu$ and $\bar{\chi} \gamma_\mu \gamma_5
\chi$ at least the divergence of the axial fermion current is
identically zero and therefore we can neglect the coupling of $\theta$
to the axial fermion current; this effect of the gauge constraint
could also be seen by shifting $B_\mu$ by $\partial_\mu \theta /\pi$.

This leads to an effective action
\bea
S_{eff} &=& \int d^2x \left( \frac{1}{2\pi} \partial_\mu
\theta (x) \partial_\mu \theta (x) -\frac{i}{2\pi} \theta
(x) \langle \epsilon_{\mu\nu} F_{\mu\nu} (x)
\rangle_{trunc} \right) \cr
&& + \int d^2x \int d^2y \frac{1}{2} \theta (x) \left\langle
\frac{1}{2\pi} \epsilon_{\mu\nu} F_{\mu\nu} (x)
\frac{1}{2\pi} \epsilon_{\mu\nu} F_{\mu\nu} (y)
\right\rangle_{trunc,c} \cr
&& +\ldots
\label{eq:2dSeff}
\eea
As mentioned, the expectation values $\langle \ldots \rangle_{trunc}$
are to be taken with respect to the truncated system described by the
Lagrangian
\beq
{\cal L}_{trunc} = \frac{1}{4e^2}F_{\mu\nu}F_{\mu\nu} +
\bar{\chi}(\slash{\partial} - i\slash{A}
-i\pi\slash{B}\gamma_5 )\chi +\bar{c}c ~~.
\label{eq:2dLtrunc}
\eeq
The manipulations leading to (\ref{eq:2dSeff}) and (\ref{eq:2dLtrunc})
did not involve any special properties of 2 dimensional field theory
and should be extendable to higher dimensions. This will be
demonstrated in the next section.

In order to investigate the effective Lagrangian further, we now have
to make use of known results in two dimensions. The easiest way to
solve the truncated theory (\ref{eq:2dLtrunc}) is to shift $B_\mu$ by
$i\epsilon_{\mu\nu} A_{\nu}/\pi$ as before. This decouples the
fermions and the (vector) gauge field $A_\mu$. Another point of view
of this decoupling was presented in connection with bosonization of
massive theories in two dimensions \cite{us}; the field
$\epsilon_{\mu\nu} B_\nu$ is essentially the same as a (vector) gauge
field like $A_\mu$, up to pure (vector) gauge degrees of freedom. The
difference to the field $A_\mu$ is the lack of a kinetic term. This
can be interpreted as an infinitely strong coupling. In two dimensions
this means total confinement; all (physical) correlation functions are
zero. The only non-vanishing expectation value is the scalar
condensate $\langle \bar{\chi}\chi\rangle_{trunc}$. This implies the
vanishing of the coupling between $A_\mu$ and the fermions.

What remains for the (vector) gauge field $A_\mu$ is the kinetic term
alone. This implies a correlation function
\beq
\left\langle
\frac{1}{2\pi} \epsilon_{\mu\nu} F_{\mu\nu} (x)
\frac{1}{2\pi} \epsilon_{\mu\nu} F_{\mu\nu} (y)
\right\rangle_{trunc,c} = \frac{e^2}{\pi^2} \delta (x-y)~~;
\label{eq:tsus}
\eeq
all higher correlation functions are vanishing. As a consequence, the
expansion in (\ref{eq:2dSeff}) terminates at second order and we are
thus back at our diagonalized version (\ref{eq:Ldiag}). The cumulant
expansion leads in this case to the exact result!

We now concentrate on the field $B_\mu$. By simple shifts we get the
following Ward identity for the truncated theory:
\beq
\langle \partial_\mu (j_\mu^5 (x) + K_\mu (x) + i B_\mu (x))\rangle =
0~.
\label{eq:WI1tr}
\eeq
Recall that (\ref{eq:2dLtrunc}) looks almost like the Lagrangian of
the original Schwinger model, except for the coupling to $B_\mu$. In
the truncated theory the current $j_\mu^5 + K_\mu$ is no longer
conserved due to the presence of precisely this topologically
non-trivial field $B_\mu$.

The consequences of this broken chiral symmetry are quite drastic.
Consider another Ward identity
\beq
\langle \partial_\mu B_\mu (x) \partial_\nu B_\nu  (y)
\rangle_{trunc,c} = \frac{1}{\pi} \partial^2 \delta (x-y) -
\left\langle
\frac{1}{2\pi} \epsilon_{\mu\nu} F_{\mu\nu} (x)
\frac{1}{2\pi} \epsilon_{\mu\nu} F_{\mu\nu} (y)
\right\rangle_{trunc,c} ~~.
\label{eq:WIB}
\eeq
Using (\ref{eq:tsus}) and recalling the purely longitudinal character
of $B_\mu$, its 2-point function in momentum space reads
\beq
\int\! d^2x\; e^{-ipx} \langle B_\mu (x) B_\nu (0) \rangle_{trunc,c} =
- f^2 \frac{p_\mu p_\nu}{p^4} (p^2 + m^2)~~,
\label{eq:WI2tr}
\eeq
implying a behaviour like $p_\mu p_\nu /p^4$ in the limit $p^2\to
0$. This behaviour can also be deduced from (\ref{eq:WIB}) after
integrating over $x$:
\beq
\int\! d^2x\; \langle \partial_\mu B_\mu (x) \partial_\nu B_\nu (y)
\rangle_{trunc} =  - \int\! d^2x\;\left\langle
\frac{1}{2\pi} \epsilon_{\mu\nu} F_{\mu\nu} (x)
\frac{1}{2\pi} \epsilon_{\mu\nu} F_{\mu\nu} (y)
\right\rangle_{trunc}
\label{eq:Vghost}
\eeq
This equation relates the low-momentum behaviour of the 2-point
function of $B_\mu$ to the topological susceptibility of the truncated
system. We already know that the latter is different from zero; this
implies a $p_\mu p_\nu /p^4$-behaviour in the limit $p^2\to 0$. Thus,
from very general arguments we find that $B_\mu$ has all the
characteristic features of the ``axial four-vector ghost'' (here axial
two-vector ghost) suggested by Veneziano \cite{Veneziano}. We believe
that one may regard $B_\mu$ as one explicit realization of this
ghost.

We would like to emphasize that the discussion of the Veneziano ghost
was based mainly on the Ward identity (\ref{eq:Vghost}), which can be
derived from chiral transformations and shifts of $B_\mu$. This
suggests a possible generalization to four dimensions. Indeed, we
shall see that the discussion concerning mass generation and the
Veneziano ghost can be formally repeated in four dimensions with the
same arguments as in the 2-dimensional case. However, it will turn out
that there are also major differences due to the much more complicated
dynamics of QCD.

\section{A flavour-singlet pseudoscalar meson from QCD}

The simplest setting with which to illustrate how our method works in
the case of QCD, is that of describing the long-distance dynamics of
the $\eta'$ meson.  Because it is a flavour singlet, the more
complicated machinery of non-Abelian chiral transformations is not
needed in this case.  We can then describe our method in simple terms,
while still having a physical situation in mind. The only complication
in this case arises from the chiral anomaly, but also this can be
taken fully into account, just as in the previous 2-dimensional
example.  The flavour-singlet states are by no means the lightest
long-distance excitations of QCD, but as shall become clear in the
following, this poses no problems either. All dynamics of the lighter
bound states will in this case still be contained in the (rotated)
quark-gluon sector.

Our starting point is in this case a generating functional for QCD (in
Euclidean space) of the form
\begin{eqnarray}
{\cal Z}_{QCD} [V,A] &=& \int\! {\cal D} [\bar{\psi},\psi]{\cal
D}\mu[G]\; e^{-\int\! d^4x\; {\cal L}_{QCD}} \cr {\cal L}_{QCD} &=&
\bar{\psi}(x) (\slash{\partial} -i\slash{G}(x)
-i\slash{V}(x) -i \slash{A} (x)\gamma_5)\psi(x) +
\frac{1}{4g^2} tr G_{\mu\nu} (x)G_{\mu\nu} (x)~.
\label{eq:ZQCD}
\end{eqnarray}
Here $V_\mu(x)$ is an external vector source and $A_\mu(x)$ an
external axial vector source, both Abelian (diagonal in the $SU(N_f)$
flavour indices). The vector potential $G_{\mu}(x)$ is the usual gluon
field, here for convenience generalized to $SU(N_c)$, and
$G_{\mu\nu}(x)$ is the corresponding field strength tensor:
\beq
G_{\mu\nu} = \partial_\mu G_\nu - \partial_\nu G_\mu -i [G_\mu
,G_\nu].
\eeq
The usual $SU(N_c)$ colour gauge symmetry of course has to be
gauge-fixed in the standard manner, including also Yang-Mills
ghosts. For the moment, we simply include these Yang-Mills
gauge-fixing terms implicitly in the gluon measure ${\cal{D}}\mu[G]$.

There is nothing unphysical implied by the coupling to external vector
and axial vector sources; these sources only serve to define
appropriate Green functions through functional differentiation. They
are clearly not intrinsically part of QCD, and will eventually be set
equal to zero.  Nevertheless, they turn out to play a rather profound
r\^{o}le in the derivation of the effective Lagrangian. They would
also, of course, acquire a physical meaning if they were to include
the couplings of the electroweak interactions.

The $\gamma$-matrices are Hermitean and obey the algebra
\beq
\{\gamma_\mu ,\gamma_\nu\} = 2\delta_{\mu\nu} \quad , \quad \{ \gamma_5
,\gamma_\mu \} =0 \quad , \quad \gamma_5 = -\gamma_1
\gamma_2 \gamma_3 \gamma_4~~.
\eeq

As in the 2-dimensional example, it is essential that the field
transformation is made only in a regularized version of the QCD
generating functional.  A convenient consistent scheme in the fermion
sector is again provided by a set of Pauli-Villars regulator
fields. It should be kept in mind that these regulators {\em only}
regularize the fermionic sector of QCD. Although we shall not consider
field transformations involving the colour gauge potentials, we of
course still need to regularize also the gluon sector of QCD. To
interpret the resulting field-transformed Lagrangian as an effective
Lagrangian with an ultraviolet cut-off $\Lambda$, a regularization
procedure with a similar cut-off ought to be introduced in the gluonic
sector. This is a non-trivial issue, but at least one correct cut-off
procedure is known to exist \cite{Warr}.

Upon the field transformation we will get two pieces, one classical
from the variation of the action in eq.(\ref{eq:ZQCD}) under the local
chiral rotation, and one quantum mechanical from the change of the
fermionic measure. The latter part can again essentially be read off
from ref.
\cite{Ball}. We rearrange it into an expansion in
decreasing powers of $\Lambda$, the ultraviolet cut-off.  Explicitly,
we can then write the regularized generating functional as
\begin{eqnarray}
{\cal Z}_\Lambda [V,A] &=& \int\! {\cal D}_\Lambda [\bar{\chi},\chi
]{\cal D}\mu[G]\; e^{-\int\! d^4x\; {\cal L}'}
\cr {\cal L}' &=& \frac{1}{4g^2}tr G_{\mu\nu}G_{\mu\nu} +
\bar{\chi} (\slash{\partial} -i\slash{G} - i\slash{V}
-i\slash{A}\gamma_5 +i\slash{\partial}\theta \gamma_5) \chi
\cr & & + {\cal{L}}_{WZ} + {\cal{L}}_J
\label{eq:Znew}
\end{eqnarray}
where the last two terms arise from the Jacobian of the
transformation.  As advertized, these last two terms can be computed
to all orders in an expansion in decreasing powers of the ultraviolet
cut-off $\Lambda$.  The first part, the Wess-Zumino term, starts at
${\cal{O}}(\Lambda^0)$:
\beq
{\cal L}_{WZ} = -\frac{iN_f}{16\pi^2}
\theta\epsilon_{\mu\nu\rho\sigma}\biggl( tr G_{\mu\nu}
G_{\rho\sigma} - 4N_c \partial_\mu V_\nu \partial_\rho V_\sigma -
\frac{4N_c}{3} \partial_\mu A_\nu \partial_\rho A_\sigma \biggr) +
{\cal O}(\Lambda^{-2}),
\label{eq:LWZ}
\eeq
while the second part -- the positive-parity part, in contrast with
the negative-parity Wess-Zumino term -- starts at
${\cal{O}}(\Lambda^2)$:
\beq
{\cal{L}}_J = \Lambda^2{\cal{L}}_2 + {\cal{L}}_0 +
\frac{1}{\Lambda^2}{\cal{L}}_{-2} + \frac{1}{\Lambda^4}
{\cal{L}}_{-4} + \ldots
\eeq
where
\bea
{\cal L}_2 &=& \frac{N_f N_c \kappa_2}{4\pi^2} \biggl( A_\mu A_\mu
-(A_{\mu} - \partial_{\mu}\theta)(A_{\mu} -
\partial_{\mu}\theta)\biggr) \cr
{\cal L}_0 &=& \frac{N_f N_c}{24\pi^2} \biggl( \partial_\mu A_\nu
\partial_\mu A_\nu - \partial_\mu(A_\nu - \partial_{\nu}\theta)
\partial_\mu (A_\nu - \partial_{\nu}\theta) \cr &&
\qquad+ 2 \Bigl( A_\mu A_\mu \Bigr)^2 - 2\Bigl((A_\mu -
\partial_\mu \theta )(A_\mu -\partial_\mu\theta)\Bigr)^2 \biggr) \cr
{\cal L}_{-2} &=& \frac{N_f \kappa_{-2}}{48\pi^2}
\biggl(N_c \partial^2 A_\mu \partial^2 A_\mu - N_c \partial^2 (A_\mu -
\partial_\mu \theta) \partial^2 (A_\mu - \partial_\mu \theta ) \cr
&& \qquad + \Bigl( A_\mu A_\mu - (A_\mu -\partial_\mu \theta )( A_\mu
-\partial_\mu \theta ) \Bigr)\;tr_c G_{\nu\rho} G_{\nu\rho} + {\cal
O}(A_\mu^4)~.
\biggr)
\label{eq:J+5}
\eea
Here, $tr_c$ denotes a trace over colour indices. In (\ref{eq:J+5}) we
list only the first three terms; the whole expansion in increasing
powers of inverse cut-off can be computed following the technique
described in ref. \cite{Ball}. Similarly, we show explicitly in
(\ref{eq:LWZ}) only the leading contribution.

The coefficients $\kappa_2$ and $\kappa_{-2}$ in eq.  (\ref{eq:J+5})
are regularization-scheme dependent constants, given here by
\beq
\kappa_n = \sum_i c_i k_i^n \ln k_i^2 \quad , \quad \sum_i c_i = 1 \quad ,
\quad \sum_i c_i k^m_i = 0 \quad \hbox{for}\quad m = 1,\ldots ,4
\eeq
where the $k_i$'s are the Pauli-Villars regulator masses in units of
the cut-off $\Lambda$, $i.e.$ $M_i = k_i\Lambda$.

This completes the part of the field-enlarging transformation in
QCD. When we next integrate over the collective fields in the path
integral, a chiral gauge symmetry again appears:
\bea
\chi (x) &\to& e^{i\alpha (x) \gamma_5} \chi (x) \cr
\bar{\chi} (x) &\to& \bar{\chi} (x) e^{i\alpha (x) \gamma_5} \cr
\theta (x) &\to& \theta (x) - \alpha (x).
\label{eq:sym5}
\eea
We shall return to the question of gauge fixing in section 3.2.

One important property of the transformed action in (\ref{eq:Znew}) is
that it already contains a kinetic energy piece for the collective
field $\theta(x)$. We can write the full Lagrangian in (\ref{eq:Znew})
in a more conventional form of a pseudoscalar field coupled to the
rotated fermion fields, the gluons and the external sources:
\bea
{\cal{L}'} &=&
\frac{1}{4g^2}tr G_{\mu\nu}G_{\mu\nu}
+ \bar{\chi} (\slash{\partial} -i\slash{G} - i\slash{V}
-i\slash{A}\gamma_5 +i\slash{\partial}\theta \gamma_5) \chi
\cr & & +\frac{N_f}{2} \partial_\mu \theta f^2 \partial_\mu \theta -
N_f A_\mu f^2 \partial_\mu \theta \cr && -\frac{N_f N_c}{12\pi^2}
\biggl(\Bigl((A_\mu -
\partial_\mu \theta )(A_\mu
-\partial_\mu \theta ) \Bigr)^2 -
\Bigl( A_\mu A_\mu \Bigr)^2 \biggr) \cr
&& - \theta \frac{iN_f}{16\pi^2} \;\epsilon_{\mu\nu\rho\sigma}
\biggl( tr_c G_{\mu\nu} G_{\rho\sigma} - 4N_c \partial_\mu V_\nu
\partial_\rho V_\sigma -\frac{4N_c}{3} \partial_\mu A_\nu
\partial_\rho A_\sigma \biggr) \cr
&& + {\cal O} (\Lambda^{-2})~.
\label{eq:L'}
\eea
The quantity $f^2$ suggesting a decay constant is actually an operator
containing higher derivatives:
\beq
f^2 = -\frac{N_c \kappa_2 \Lambda^2}{2\pi^2} + \frac{N_c}{12\pi^2}
\partial^2 - \frac{N_c \kappa_{-2}}{24\pi^2 \Lambda^2} \partial^2
\partial^2 - \frac{\kappa_{-2}}{24 \pi^2
\Lambda^2} \; tr_c  G_{\nu\rho} G_{\nu\rho} + \ldots
\label{eq:f2}
\eeq
The dots denote higher-order gluonic terms, derivatives and
combinations of them divided by suitable powers of $\Lambda$. As such,
$f^2$ summarizes all contributions of second order in $A_\mu$ in the
chiral Jacobian.

\subsection{{\sc{Induced Regularization}}}

It is interesting to note that not only have we automatically
generated, after a trivial rescaling of $\theta(x)$, a standard
canonical kinetic energy term for this new field\footnote{It seems
that the kinetic term has the wrong sign. However, we are dealing with
a regularized theory and thus with {\em bare} parameters; there is no
reason why these parameters should have a definite sign.  Indeed, the
number $\kappa_2$ is completely scheme dependent and may have an
arbitrary sign.}, we actually have a {\em Pauli-Villars regularized}
version of it.  This regularization is not put in by hand, but induced
by the chiral rotation, and its associated regularized Jacobian.  To
see this, let us concentrate on the terms in (\ref{eq:L'}) that are
quadratic in $\theta$, and let us furthermore neglect the gluonic
terms in the expression (\ref{eq:f2}) for $f^2$. Suppose we introduce
a pseudoscalar field $\eta_0$ with the canonical dimension of $mass$
as
\beq
\theta = \frac{1}{\sqrt{N_f}f_0}\eta_0
\eeq
and assume that $f_0$ can be viewed as a bare coupling.
Dimensionally, this coupling also goes as $[mass]^1$, and must hence
be proportional to the ultraviolet cut-off $\Lambda$, the only scale
in this regularized theory. Indeed, identifying $f_0^2$ with the
leading term in (\ref{eq:f2}),
\beq
f_0^2 = -\frac{N_c \kappa_2 \Lambda^2}{2\pi^2}
\eeq
leads to a Lagrangian typical for a pseudoscalar field:
\beq
{\cal L}_{\eta_0} = \frac{1}{2} \partial_\mu \eta_0 \partial_\mu
\eta_0 - A_\mu \sqrt{N_f} f_0 \partial_\mu \eta_0 + \ldots
\eeq
The dots denote higher derivative terms, gluonic terms and self
interactions of the pseudoscalar field. Note that $f_0$ is both
cut-off dependent and scheme dependent.

Let us now look at the higher derivative terms. In a perturbative
sense, the propagator for the field $\eta_0$ can be derived from the
bilinear part of ${\cal L}'$; keeping also the first of the higher
derivative terms of $f^2$ in (\ref{eq:f2}), it would be, in a momentum
representation,
\beq
G(p^2) = \frac{1}{p^2} - \frac{1}{p^2 + 6\kappa_2\Lambda^2}~.
\eeq
This is a Pauli-Villars regularized bosonic propagator with the
regulator mass proportional to $\Lambda^2$, the fermionic cutoff
parameter. Actually, in four dimensions we will encounter both
quadratic and logarithmic divergences, and one Pauli-Villars regulator
therefore does not suffice to regularize the theory completely. This
extra Pauli-Villars regulator mass for $\theta$ is {\em also}
automatically provided by the chiral rotation. Going to one order
higher in the expansion of the free part of $f^2$, we find a term
which in momentum space is of the form $p^4/\Lambda^4$. The inclusion
of this term is equivalent to the addition of one more Pauli-Villars
regulator field in the $\theta$-sector.

One may well ask why we stop the $\Lambda^{-1}$-expansion at this
point, since it seems to lead to such drastic consequences as
introducing two Pauli-Villars regulator masses to order
${\cal{O}}(\Lambda^{-2})$. What if we considered yet one more term in
the expansion, would that not significantly alter the results? The
answer is that the two Pauli-Villars regulators we have included are
already sufficient to regularize the ultraviolet divergences
associated with the $\theta$-dynamics (at least in perturbation
theory). We could of course add yet one higher order if we wished, and
also include it in the kinetic energy term instead of treating it as a
perturbation.  This corresponds to an oversubtracted Pauli-Villars
regularization, a valid but inconvenient regularization scheme if we
send the cut-off $\Lambda$ to infinity in the end.  Then the
$\theta$-propagator with just these two Pauli-Villars regulator terms
included is enough to regularize the theory. All higher-order terms
are then not just formally, but truly (at least within perturbation
theory) suppressed by extra powers of $\Lambda^{-1}$. Of course, if we
do not wish to send the cut-off $\Lambda$ to infinity, then all these
extra terms in $f^2$ must be kept for consistency.

The same phenomenon of induced regularization for the collective field
occurs in the (1+1)-dimensional case as well \cite{us}, and there is
probably a simple intuitive reason for it. The point is that we are
throughout performing field transformations within a {\em regularized}
fermionic path integral: all ultraviolet divergences have explicitly
been removed. Even after a series of field-enlarging transformations
(and the required gauge fixing of the new local symmetry) of such a
form that we end up with new propagating fields, the generating
functional must still be ultraviolet regularized. The fermionic
Jacobian precisely reflects this fact by inducing appropriate
regulator terms for the new fields where short-distance singularities
threaten to appear.

The above discussion has not covered the gluonic contributions to
$f^2$ in (\ref{eq:f2}). If these were external fields, then we can
deduce from the arguments given above that these terms are suppressed
by inverse powers of the cut-off. What changes if these are dynamical
fields, as in QCD? In this case we know that there may occur gluon
condensates like $\langle G^2 \rangle$.  In regularized QCD the latter
is necessarily of order $\Lambda^4$ and will thus contribute to the
non-derivative part of $f^2$. Simple dimensional reasoning leads to
the conclusion that there are similar contributions to the derivative
terms in $f^2$.  In principle, the higher-order gluon condensates have
to be included in the same manner. All these terms are calculable from
the $1/\Lambda$-expansion of the chiral Jacobian. We will come back to
these gluonic terms in section 3.3.

\subsection{{\sc An Anomalous Gauge Fixing in Four Dimensions}}

Useful forms of the effective Lagrangian are derived by judicious
choices of gauge fixing. Whereas the gauge $\theta(x) = 0$ trivially
gives us back cut-off QCD in its original formulation, essentially all
other gauge choices that remove part of the QCD degrees of freedom
will lead to non-trivial effective Lagrangians. We shall here use some
of the lessons learned in the previous 2-dimensional example, and
consider the expectation value of the axial singlet current as given
by a functional derivative with respect to $A_\mu$ at $A_\mu =0$. It
reads
\beq
i\langle \bar{\psi} \gamma_\mu \gamma_5 \psi \rangle = i\langle
\bar{\chi} \gamma_\mu \gamma_5 \chi \rangle + N_f f^2
\partial_\mu \theta + \ldots
\label{eq:trAxCur}
\eeq
The additional terms represented by dots are at least of third order
in $\theta(x)$. The whole expression is, of course, gauge invariant;
however, the individual components on the r.h.s. are gauge
dependent. Expectation values in eq.  (\ref{eq:trAxCur}) are with
respect to the fermionic fields only.

In order to keep the discussion of gauge fixing as simple as possible
at this point, let us first ignore all ${\cal{O}}(\Lambda^{-2})$ (and
higher) corrections in the operator $f^2$, which then reduces to
$f_0^2$ as defined in the last subsection.  The corrections neglected
include non-trivial gluonic contributions, whose significance we shall
discuss later.

A comparison with the solvable two-dimensional case
\cite{us} is again useful at this point. There, the equivalent of
eq.(\ref{eq:trAxCur}) for the same chiral transformation contained
only a term linear in $\theta$. By gauge fixing on a suitable linear
combination of the resulting two terms comprising the divergence of
the physical gauge-invariant axial current, it is possible to find
\cite{us} the (presumably unique) gauge-fixing function that switches
smoothly between fermionic and bosonic formulations. Since we at least
partially wish to achieve something close to ``bosonization" of the
low-energy QCD Lagrangian, let us first try to choose a gauge-fixing
function $\Phi$ analogous to the (1+1)-dimensional case:
\beq
\Phi =  \Delta i\frac{\partial_\mu}{N_f f_0^2\partial^2}
\bar{\chi} \gamma_\mu \gamma_5\chi + (1-\Delta)  \theta~.
\label{eq:Phi'}
\eeq
This choice of gauge follows the same philosophy as in (1+1)
dimensions: the $\theta$-dependent part of the divergence of the axial
current should describe a fraction $\Delta$ of the divergence of the
{\em physical} axial current expressed through the {\em original}
fermionic degrees of freedom $\bar{\psi}$ and $\psi$.

The above gauge-fixing function is, however, defined in a very formal
manner on account of the inverse Laplacian.  Furthermore, if we wish
to implement $\Phi$ as a delta-function constraint in the path
integral, we should be careful that it satisfies correct
transformation properties.  Under a global chiral rotation,
\beq
\bar{\chi}(x) \to \bar{\chi}(x)e^{i\alpha\gamma_5} ~~,~~~~~
\chi(x) \to e^{i\alpha\gamma_5}\chi(x) ~,
\eeq
the object involving $\bar{\chi}\gamma_{\mu}\gamma_5\chi$ remains {\em
classically} invariant but quantum mechanically it shifts due to the
chiral anomaly. We find the shift from the expansion given in
eq. (\ref{eq:L'}):
\beq
i\frac{\partial_\mu}{N_f f_0^2\partial^2}
\bar{\chi} \gamma_\mu \gamma_5\chi \to
i\frac{\partial_\mu}{N_f f_0^2\partial^2}
\bar{\chi} \gamma_\mu \gamma_5\chi + \alpha ~.
\eeq
The formal operator $N_f f_0^2\partial^2$ is included precisely in
order to cancel the additional factors multiplying $\alpha$.

The action does not remain invariant under constant chiral rotations,
but shifts due to the axial anomaly:
\beq
S' = \int\! d^4x\; {\cal{L}}' \to S' - 2i N_f \alpha \int\! d^4x
\frac{1}{32\pi^2}\; \epsilon_{\mu\nu\rho\sigma} tr_c G_{\mu\nu}
G_{\rho\sigma} ~.
\eeq
Assuming that we sum only over integer winding numbers, the action
does, however, remain invariant under constant chiral rotations of the
form $\alpha = n\pi/N_f$. This means that $\theta(x)$ is only globally
defined modulo $\pi/N_f$.

Gauge-fixing the constant $\theta(x)$ modes is then again
non-trivial. The gauge-fixing constraint must respect the above
periodicity property; there must, even in the gauge-fixed path
integral, be no distinction between $\theta(x)$ and $\theta(x) +
n\pi/N_f$. If we choose a $\delta$-function constraint to implement
the gauge choice, this $\delta$-function must then necessarily be {\em
globally periodic}, quite analogous to the 2-dimensional case.

We can again represent such a globally periodic $\delta$-function by
means of a Nakanishi-Lautrup auxiliary field $b(x)$, so that this
gauge-fixing function just provides a few new terms in the
action. However, these new terms in the action will in general modify
the relevant chiral Jacobian, and for consistency -- or, equivalently,
if we wish to keep BRST symmetry in the gauge fixing -- new terms must
be added to the action to compensate for the change in the
Jacobian. This leads to a rather involved procedure of gauge fixing,
and we have preferred to present a simpler derivation. The final
result agrees with that obtained through the direct method.

This is based on the fact that quantum mechanically we can consider
the above gauge-fixing function as derived from a constraint in the
{\em original} representation of QCD of the form
\beq
\Phi' =  \Delta i\frac{\partial_\mu}{N_f f_0^2\partial^2}
\bar{\psi} \gamma_\mu \gamma_5\psi + \theta ~.
\eeq
We shall here illustrate the gauge-fixing procedure by this shortcut.
The equivalence of $\Phi$ and $\Phi'$ is valid at the level of
expectation values. In the final gauge-fixed action there is of course
an enormous amount of freedom left, since we can add any term
proportional to only $b(x)$ (and higher powers thereof), as long as
these terms do not multiply any functions of $\theta$. These terms do
not affect physical gauge-invariant Green functions. We shall make use
of this freedom shortly.

As mentioned, it is convenient to implement the periodic
$\delta$-function by a functional Fourier representation. If we
introduce the $\delta$-function constraint in the shortcut manner
described above, this corresponds to
\bea
\delta(\Phi') &=& \int\! {\cal{D}}[b]\; \exp\left[-\int\! d^4x d^4z\; b(x)
\left(\frac{i\Delta}{N_f f_0^2}\partial^{-2}_{(x-z)}
\partial_{\mu}^{(z)}J^5_{\mu}(z) - \delta(x-z)\theta(z)\right)
\right] \cr
&=& \int\! {\cal{D}}[b]\; \exp\left[-\int\!  d^4x\;
\left(\frac{i\Delta} {N_f f_0^2}B_{\mu}(x)J^5_{\mu}(x) +
\theta(x)\partial_{\mu}B_{\mu}(x)
\right)\right] ~,
\eea
where the axial vector field $B_{\mu}(x)$ is defined by
\beq
B_{\mu}(x) \equiv \int\! d^4y d^4z\; b(y)\partial^{-2}_{(y-z)}
\partial_{\mu}^{(z)}\delta(z-x) ~.
\eeq
Note that this implies $b(x) = -\partial_{\mu}B_{\mu}(x)$ as in the
2-dimensional case.  Global periodicity of the $\delta$-function means
that $b$ is constrained,
\beq
\int\! d^4x\; b(x) = -\int\! d^4x\; \partial_{\mu}B_{\mu} = ik N_f ~,
\eeq
where $k$ is an arbitrary integer. This global constraint means that
$b(x)$ (or $\partial_{\mu}B_{\mu}$) share certain properties with
topologically non-trivial fields. But we of course have no direct
interpretation in terms of gauge potentials. In two dimensions, the
same global constraint has a more physical interpretation in terms of
an Abelian gauge field of non-trivial topology
\cite{us}. Nevertheless, these topological properties of $b(x)$ play a
similarly profound r\^ole in chiral Ward identities of a truncated QCD
(to be defined in a precise manner later) as in the 2-dimensional
case.

With this implementation of the gauge fixing in the original
Lagrangian, we can now perform the same chiral transformation as
before.  The only change is that in the transformed Lagrangian
${\cal{L}}'$, the external axial vector source $A_\mu$ has to be
replaced by $A_\mu -\frac{\Delta}{N_f f_0^2} B_\mu$ and a term $\theta
\partial_\mu B_\mu$ has to be added.

After gauge-fixing the only remnant of the $U(1)$ axial gauge symmetry
is the BRST symmetry
\begin{eqnarray}
\delta \bar{\chi}(x) &~=~& i\bar{\chi}(x)\gamma_5c(x) \cr
\delta \chi(x) &~=~& -ic(x)\gamma_5\chi(x) \cr
\delta \theta(x) &~=~& -c(x) \cr
\delta c(x) &~=~& 0 \cr
\delta \bar{c}(x) &~=~& b(x) \cr
\delta b(x) &~=~& 0.
\end{eqnarray}
The ghost term is trivial in the present case:
\beq
{\cal{L}}_{ghost} = \bar{c}c
\eeq
It decouples and will be neglected in the following.

The complete gauge-fixed Lagrangian now reads
\bea
{\cal L}'' &=& \bar{\chi} \biggl( \slash{\partial} -i\slash{G}
-i\slash{V} -i\Bigl(\slash{A} -
\frac{\Delta}{N_f f_0^2}\slash{B} -\slash{\partial} \theta
\Bigr) \gamma_5\biggr) \chi + \bar{c} c + {\cal L}_{YM}  \cr
&& +\frac{N_f f_0^2}{2} \partial_\mu \theta \partial_\mu \theta - N_f
f_0^2 A_\mu \partial_\mu \theta - (1-\Delta) \theta \partial_\mu B_\mu
\cr && -\frac{N_f N_c}{12\pi^2} \biggl(\Bigl((A_\mu -\frac{\Delta}{N_f
f_0^2} B_\mu - \partial_\mu \theta )(A_\mu -\frac{\Delta}{N_f f_0^2}
B_\mu -\partial_\mu \theta ) \Bigr)^2 \cr &&- \Bigl( (A_\mu
-\frac{\Delta}{N_f f_0^2} B_\mu )(A_\mu -\frac{\Delta}{N_f f_0^2}
B_\mu) \Bigr)^2 \biggr) \cr && -
\theta \frac{iN_f}{16\pi^2} \;\epsilon_{\mu\nu\rho\sigma}
\biggl( tr_c  G_{\mu\nu} G_{\rho\sigma} - 4 N_c \partial_\mu V_\nu
\partial_\rho V_\sigma -\frac{4N_c}{3} \partial_\mu A_\nu \partial_\rho
A_\sigma \biggr) \cr && + {\cal O} (\Lambda^{-2})~.
\label{eq:Lgf5}
\eea
Due to its longitudinal character $B_\mu$ does not appear in the
WZ-like term of the last line.

It is convenient to choose a slightly different gauge which will only
affect higher order correlation functions of our gauge fixing
expression $\Phi'$. As already mentioned we are allowed to add
arbitrary terms containing $b(x)$ or powers thereof as long as they
are not coupled to $\theta (x)$. To be precise, we choose them such
that the gauge fixed Lagrangian looks like
\bea
{\cal L}'' &=& \bar{\chi} \biggl( \slash{\partial} -i\slash{G}
-i\slash{V} -i\Bigl(\slash{A} -
\frac{\Delta}{N_f f_0^2}\slash{B} -\slash{\partial} \theta
\Bigr) \gamma_5\biggr) \chi + \bar{c} c + {\cal L}_{YM}  \cr
&& +\frac{N_f f_0^2}{2} \partial_\mu \theta
\partial_\mu \theta - N_f f_0^2 A_\mu \partial_\mu \theta - (1-\Delta) \theta
\partial_\mu B_\mu \cr
&& -\frac{N_f N_c}{12\pi^2} \biggl(\Bigl((A_\mu -\frac{\Delta}{N_f
f_0^2} B_\mu - \partial_\mu \theta )(A_\mu -\frac{\Delta}{N_f f_0^2}
B_\mu -\partial_\mu \theta ) \Bigr)^2  - \Bigl( A_\mu A_\mu \Bigr)^2
\biggr) \cr && - \theta \frac{iN_f}{16\pi^2}
\;\epsilon_{\mu\nu\rho\sigma} \biggl( tr_c G_{\mu\nu} G_{\rho\sigma} -
4 N_c \partial_\mu V_\nu \partial_\rho V_\sigma -\frac{4N_c}{3}
\partial_\mu A_\nu \partial_\rho A_\sigma \biggr) \cr && + {\cal O}
(\Lambda^{-2})
\label{eq:Lgf6}
\eea

\subsection{{\sc{The $\Delta = 1$ gauge}}}

In two space-time dimensions, the case $\Delta=1$ corresponds to the
``bosonization gauge" \cite{us}. It is therefore natural to consider
the corresponding analogue here in four dimensions. Of course, we have
no hope of completely bosonizing QCD, but we should not be surprised
to find that the corresponding $\Delta=1$ gauge in four dimensions
most conveniently extracts the pseudoscalar collective degrees of
freedom. Actually, we can only bosonize those degrees of freedom
representing the longitudinal component of $J_{5\mu}$. With $\Delta
=1$, ${\cal L}''$ becomes
\bea
{\cal L}'' &=& \bar{\chi} \biggl( \slash{\partial} -i\slash{G}
-i\slash{V} -i\Bigl(\slash{A} -
\frac{1}{N_f f_0^2}\slash{B} -\slash{\partial} \theta
\Bigr) \gamma_5\biggr) \chi + \bar{c} c + {\cal L}_{YM}  \cr
&& +\frac{N_f f_0^2}{2} \partial_\mu \theta
\partial_\mu \theta - N_f f_0^2 A_\mu \partial_\mu \theta \cr
&& -\frac{N_f N_c}{12\pi^2} \biggl(\Bigl((A_\mu -\frac{1}{N_f f_0^2}
B_\mu - \partial_\mu \theta )(A_\mu -\frac{1}{N_f f_0^2} B_\mu
-\partial_\mu \theta ) \Bigr)^2 - \Bigl( A_\mu A_\mu
\Bigr)^2 \biggr) \cr
&& - \theta \frac{iN_f}{16\pi^2} \;\epsilon_{\mu\nu\rho\sigma} \biggl(
tr_c G_{\mu\nu} G_{\rho\sigma} - 4N_c \partial_\mu V_\nu \partial_\rho
V_\sigma -\frac{4N_c}{3} \partial_\mu A_\nu \partial_\rho A_\sigma
\biggr) \cr && + {\cal O} (\Lambda^{-2})~.
\label{eq:LDel1}
\eea
As expected, this involves all fields -- fermionic and bosonic --
interacting in all respects.  We are nowhere near a bosonized form of
the QCD Lagrangian. But, as we shall see, many of the simplifying
features of two-dimensional bosonization nevertheless remain hidden in
this form of the Lagrangian.

In order to view (\ref{eq:LDel1}) as an effective Lagrangian, we need
additional input. The obvious choice would be to identify the
$\theta$-field with the flavour-singlet pseudoscalar field of the
$\eta'$ meson, in appropriate units. But depending on the questions we
wish to answer, such an explicit identification may not be
necessary. Certainly, eq.  (\ref{eq:LDel1}) gives in the $\Lambda \to
\infty$ limit the correct QCD action for describing the dynamics of the
composite operator $J^5_{\mu}(x) =
i\bar{\psi}\gamma_{\mu}\gamma_5\psi(x)$ of the original quark fields
supposing the limit for the gluonic terms of the chiral Jacobian is
taken in an appropriate way. Taking one partial derivative, we can
equally well describe $\partial_{\mu}J^5_{\mu}(x)$, which is a
non-zero operator due to the chiral anomaly. It has quantum numbers
$J^{PC} = 0^{-+}$, and is a singlet under flavour.  As such, this
object should have a non-vanishing overlap with the physical $\eta'$
meson. For example, if we were able to compute the long-distance
fall-off of the corresponding two-point correlation function, this
should provide us with the mass of the lowest-lying state of these
quantum numbers. Ignoring the possibility of a lighter glueball with
the same quantum numbers, this is the mass of the $\eta'$ meson.

Let us now proceed with such a calculation. Going back to the defining
equation (\ref{eq:ZQCD}), we note that the connected 2-point function
of $\partial_{\mu}J^5_{\mu}(x)$ can be obtained by differentiating
twice with respect to a pseudoscalar source $\sigma(x)$ defined by
splitting $A_{\mu} = \partial_{\mu}\sigma(x) + A_\mu^T$ into a
longitudinal and a transverse part. Shifting $B_\mu$
\beq
B_\mu (x) \to B_\mu (x) + N_f f_0^2 \partial_\mu \theta(x) - N_f f_0^2
\partial_\mu
\sigma (x)
\label{eq:shift}
\eeq
leads to a Lagrangian
\begin{eqnarray}
{\cal L}''' &=& \bar{\chi}\biggl(\slash{\partial} -i\slash{G}
-i\Bigl(\slash{A}^T - \frac{1}{N_f f_0^2}\slash{B}
\Bigr)\gamma_5 \biggr)
\chi + {\cal L}_{ghost} + {\cal L}_{YM}\cr
&& -\frac{N_f N_c}{12\pi^2} \biggl(\bigl((A_\mu^T - \frac{1}{N_f
f_0^2} B_\mu ) (A_\mu^T -\frac{1}{N_f f_0^2} B_\mu ) \bigr)^2 - \bigl(
A_\mu A_\mu \bigr)^2
\biggr) \cr
&& +\frac{N_f f_0^2}{2}
\partial_\mu \theta \partial_\mu \theta - N_f f_0^2 \partial_\mu \sigma
\partial_\mu \theta \cr
&& - \theta \frac{iN_f}{16\pi^2} \;\epsilon_{\mu\nu\rho\sigma}
\biggl( tr_c G_{\mu\nu} G_{\rho\sigma} - 4 N_c\partial_\mu
V_\nu \partial_\rho V_\sigma -\frac{4N_c}{3} \partial_\mu A^T_\nu
\partial_\rho A^T_\sigma \biggr) \cr
&& + {\cal O} (\Lambda^{-2})
\label{eq:L'''}
\end{eqnarray}
Apart from contact terms only a linear coupling of $\sigma$ to
$\theta$ is left. The remaining part of ${\cal O} (\Lambda^{-2})$ is
also independent of $\theta$ because it contains $\theta$ only in the
combination $B_\mu + N_f f_0^2 \partial_\mu \theta$; after the shift
(\ref{eq:shift}) $\theta$ dissappears from these terms.

We can now derive some exact Ward identities, setting the external
sources to zero: The original anomalous Ward identity
\beq
\partial_\mu \langle J_{5\mu} \rangle = i\partial_\mu \langle \bar{\psi}
\gamma_\mu \gamma_5 \psi \rangle = -i \frac{N_f}{16\pi^2} \langle G\tilde{G}
\rangle + {\cal O}(\Lambda^{-2})
\eeq
is now just the equation of motion for the field $\theta$:
\beq
f_0^2 \partial^2 \langle \theta \rangle = -i
\frac{1}{16\pi^2}
\langle G\tilde{G} \rangle +{\cal O} (\Lambda^{-2})
\eeq
with the shorthand notation
\beq
G\tilde{G} (x) = \epsilon_{\mu\nu\rho\sigma} tr_c G_{\mu\nu} (x)
G_{\rho\sigma} (x)
\label{eq:defGD}
\eeq

Analogously we find an anomalous expression for the 2-point function
in the original QCD representation,
\bea
\langle \partial_\mu J_{5\mu} (x) \partial_\nu J_{5\nu} (y) \rangle &=& -
\left(\frac{N_f}{16\pi^2} \right)^2 \langle G\tilde{G} (x) G\tilde{G} (y)
\rangle \cr
&& - N_f f_0^2 \partial^2\delta (x-y) +{\cal O} (\Lambda^{-2})~~.
\label{eq:chWI2}
\eea
Note the presence of a contact term in eq. (\ref{eq:chWI2}). We will
come back to it at the end of this subsection.  The same identity can
be derived considering a simple infinitesimal shift of $\theta$ to
second order:
\bea
f_0^4 \langle \partial^2 \theta (x) \partial^2\theta (y))
\rangle &=& - \left( \frac{1}{16\pi^2} \right)^2 \langle
G\tilde{G} (x) G\tilde{G} (y) \rangle \cr & & - f_0^2 \partial^2
\delta (x-y) +{\cal O}(\Lambda^{-2})
\label{eq:2pt}
\eea
These two examples illustrate that for appropriate Green functions our
$\Delta=1$ gauge really identifies
\beq
\partial_\mu J_{5\mu} \sim N_f f_0^2 \partial^2 \theta
\eeq
as {\em operators}. Actually, this identification holds only up to
fourth order correlation functions. The reason is that -- as already
mentioned -- we have modified our original gauge by the inclusion of
terms depending on $b(x)$ (resp.  $B_\mu (x)$). These terms were
precisely of fourth and higher order.

The gauge-fixing procedure presented above can therefore be understood
from another point of view. Essentially it amounts to introducing
explicitly, at the Lagrangian level, an ``interpolating" field
according to the relation
\beq
\partial_\mu J_{5\mu} = \sqrt{N_f} f_0 \partial^2\eta_0~,
\eeq
where $f_0$ is the corresponding decay constant. On mass shell, $p^2 =
m_{\eta_0}^2$, this would indeed be a conventional definition of an
interpolating field.  In our gauge, however, the interpolating field
is identified even off-shell, as it should be in a path integral
framework.

Note that we are not claiming, and need not claim, that the field
$\eta_0(x)$ is to be identified with the $\eta'$ meson. Only at large
distances, where other isosinglet pseudoscalar states are suppressed
because of their presumed larger mass, can we indirectly make such an
identification, as in the usual reduction formalism.  The present
gauge choice simply enforces that a portion $\Delta$ (and in this
particular case of $\Delta = 1$, {\em all}) of the divergence of the
physical axial current $J_{5\mu} = i\bar{\psi} \gamma_\mu
\gamma_5 \psi $ is carried by the collective field.

We now formally integrate out all fields in (\ref{eq:L'''}) except
$\theta$ to arrive at an effective Lagrangian
\beq
{\cal L}_{eff} = \frac{F^2_0}{2} \partial_\mu \theta
\partial_\mu \theta +
\frac{F_0^2 M_0^2}{2} \theta^2 + \ldots
\eeq
The dots denote higher derivative terms and self-interactions of order
$\theta^3$. The parameters $F_0$ and $M_0$ are defined through
\beq
F_0^2 M_0^2 = \int\! d^4x\; \left\langle \frac{N_f}{16\pi^2}
G\tilde{G} (x)
\frac{N_f}{16\pi^2} G\tilde{G} (0) \right\rangle_{trunc}
\label{eq:M0}
\eeq
and
\beq
F_0^2 = N_f f_0^2 - \int\! d^4x\; \frac{x^2}{8} \left\langle
\frac{N_f}{16\pi^2} G\tilde{G} (x) \frac{N_f}{16\pi^2} G\tilde{G} (0)
\right\rangle_{trunc}~~.
\label{eq:F0}
\eeq

The parameters $F_0,M_0$ have to be interpreted as {\em bare}
ones. The expression for $F_0^2 M_0^2$ in ({\ref{eq:M0}) is formally
the same as derived by Witten
\cite{Witten} and Veneziano \cite{Veneziano} in the limit
$N_c \to
\infty$. In our case this relation is valid for arbitrary $N_c$. But
the expectation values $\langle \ldots \rangle_{trunc}$ have to be
taken with respect to a ``truncated'' version of QCD:
\bea
{\cal L}_{trunc} &=& \bar{\chi}\biggl(\slash{\partial} -i\slash{G}
+i\frac{1}{N_f f_0^2} \slash{B}\gamma_5 \biggr)\chi + {\cal L}_{ghost}
+ {\cal L}_{YM}\cr && -\frac{ N_c}{12\pi^2 N_f^3 f_0^8} \biggl( B_\mu
B_\mu \bigr)^2 \cr && + {\cal O} (\Lambda^{-2})~,
\eea
where we have neglected external sources.

At first glance one would argue that the topological susceptibility
has to be zero in such a theory because of the massless quarks. If
this were true, then our field $\theta$ would be massless. Indeed, for
the full theory we can derive from the Ward identities (\ref{eq:2pt})
that
\beq
\int\! d^4x\; \langle G\tilde{G} (x) G\tilde{G} (0) \rangle = 0~.
\label{eq:mom1}
\eeq
This is a well-known result in massless QCD. Let us do an analogous
step for the truncated theory. In particular we get from a relation
resembling (\ref{eq:2pt}) the identity
\bea
\int\! d^4x\; \langle \partial_\mu B_\mu (x) \; \partial_\nu B_\nu (0)
\rangle_{trunc} &=& - \biggl(\frac{N_f}{16\pi^2}\biggr)^2 \int\!
d^4x\; \langle G\tilde{G} (x) G\tilde{G} (0)\rangle_{trunc} \cr && +
{\cal O} (\Lambda^{-2})~~,
\eea
the analogue of relation (50) in two dimensions.  Now we see the
importance of gauge-fixing the zero-modes in a proper manner: it
forced us to introduce the non-trivial constraint $\int\! d^4x\;
\partial_\mu B_\mu (x) = const.$. It is precisely this constraint that
may imply a non-vanishing topological susceptibility in the truncated
theory and thus a nonvanishing mass for the field $\theta$. Without
this constraint, one could decouple the $B_{\mu}$-field from the
fermions through a chiral rotation. The standard proof of a vanishing
topological susceptibility in the theory with massless quarks would
then go through. However, when the $B_{\mu}$-field is constrained in
the manner shown, this decoupling by means of a chiral rotation is
impossible \cite{Bardakci}, and the topological susceptibility may be
non-vanishing even (as in this truncated theory) when the fermions are
massless. The field $B_\mu$ is probably the closest one can get at an
explicit QCD Lagrangian realization of the Veneziano ghost
\cite{Veneziano}.

The essence of this exercise was to get some insight into the details
of the gauge-fixing mechanism. Let us now try to understand these
results from a more phenomenological point of view. For that purpose
it is useful to compare with the argumentation of Witten in
ref.\cite{Witten}, where it was argued that the vanishing of the
topological susceptibility in massless QCD can be explained as a
cancellation of the pure gluonic part by a contribution from the
$\eta'$ meson.  In our case, the appearance of a non-zero topological
susceptibility in the {\em truncated} QCD would be interpreted as an
effect of the $b$-field interaction.  This additional interaction,
which arises through the gauge constraint, can be thought of as
removing the $\eta'$-part of the topological susceptibility.  The
picture of the mass generation of the $\eta'$ for finite $N_c$ that
emerges then seems to coincide with the one of ref. \cite{Dow}. In
fact, we see here one explicit Lagrangian realization of what is meant
by the notion of ``no $\eta'$ insertion", which plays such a crucial
r\^{o}le in ref.\cite{Dow}. In our formulation this subtraction of the
$\eta'$ degree of freedom from the QCD generating functional is done
in an exact manner; the expectation values with no $\eta'$ insertions
simply mean expectation values taken with respect to the {\em chirally
rotated} quark fields $\bar{\chi},\chi$ and the gluons.

At the moment, we have not specified how to renormalize the effective
theory. A major difficulty with the present approach is that
everything is expressed in terms of bare parameters in the cut-off
theory. Being explicitly cut-off dependent, we are not surprised to
find that the coefficients of the effective couplings are also scheme
dependent. The whole set of effective one-loop interactions between
the bosonic collective fields and left-over QCD degrees of freedom
indeed follow directly from the Pauli-Villars regulator fields.  This
is just as in the solvable case of two dimensions \cite{us}.

In principle, the renormalization prescription should be equivalent to
the one for QCD in its original representation. The situation is,
however, further complicated by the fact that we are necessarily
dealing with a generating functional of composite operators.  One
renormalization procedure including arbitrary insertions of such
operators has recently been suggested in ref.\cite{Shore} (see also
ref. \cite{ShVen}).  As it stands, the unrenormalized theory has, with
massless quarks, only one mass scale: that of the cut-off $\Lambda$.
This means that all dimensionful couplings in the effective theory are
given by powers of this ultraviolet cut-off. In the renormalized
theory this cut-off becomes replaced by a physical mass scale, to be
extracted from experiments. In the end, if one integrates out all
gluonic and quark degrees of freedom and leaves only the collective
fields, the physical couplings are directly related to gluonic and
fermionic correlators, moments thereof, and condensates. The precise
relationship between the couplings of the collective field Lagrangian
and these vacuum expectation values will be of roughly the kind
discussed in the case of the Witten-Veneziano relation above, but will
of course require a non-perturbative treatment. The fact that physical
couplings will be related to these Green functions is also evident if
we return to the definition of the gauge-fixing condition
(\ref{eq:Phi'}) in subsection 3.2..  The term containing $f_0^2$
should in fact depend on $f^2$ with contributions also from gluonic
fields, and a full treatment should incorporate the effect of
integrating out the gluonic degrees of freedom. Intuitively, we would
expect that one major effect of such a renormalization program would
relate $f^2$ to gluonic condensates.

An important point needs to be mentioned here. If we turn off all
gluonic interactions, we can still go through the steps of deriving
the ``effective Lagrangian'' by means of chiral rotations in the
cut-off theory. It appears as if the only difference between these two
effective Lagrangians arises from ``small gluonic
corrections''. Ignoring complications due to the anomalous gauge
fixing, this would seem to imply that a convenient effective
Lagrangian for a theory of {\em free} fermions is of essentially the
same form as the one derived here for QCD! The same argument can,
incidentally, be raised against the purely bosonic effective
Lagrangians of refs. \cite{chiral,Espriu}. What is the resolution of
this apparent contradiction? As we have emphasized earlier, the
collective field technique is not tied to such notions as spontaneous
symmetry breaking and the existence of (pseudo-) Goldstone bosons. We
can {\em always} introduce given collective fields in a theory,
independently of whether these collective fields may be related to
asymptotic states of the theory. For a free theory, a collective field
representation is valid, but of limited use: The corresponding
collective field mode will decay into the free constituents.
Alternatively, we can look at this from the point of view of
renormalized couplings in the effective Lagrangian. For a free theory,
the only scale with which to specify effective couplings will remain
the (unphysical) ultraviolet cut-off. In the full theory of QCD, these
effective couplings will be specified by dynamics, including such
crucial features as chiral symmetry breaking, gluonic condensates
etc.  What this shows very explicitly is that the term ``gluonic
corrections'' may be quite misleading; in the end a large fraction of
the effective couplings of the resulting chiral Lagrangian for the
collective field may be given by the values of gluonic condensates,
moments of such condensates, the chiral condensate, and so on.

Note that in the limit $\Lambda \to \infty$, the generating functional
with the Lagrangian in eq. (\ref{eq:LDel1}) is an {\em exact} rewrite
of the generating functional of QCD, eq.  (\ref{eq:ZQCD}). No
approximations are involved at this point as long as we perform the
limit $\Lambda \to \infty$ for the whole expansion of the chiral
Jacobian in a careful manner, taking into account possible gluonic
condensates. Even for finite $\Lambda$ can we, in principle, make the
representation as accurate as we wish by including a sufficient number
of terms in the known $1/\Lambda$ expansion. In this sense, the
rewrite of the generating functional is always exact, for any value of
$\Lambda$.

We emphasize that a non-perturbative derivation of the physical
couplings is possible in principle, since the cut-off effective
Lagrangian contains the same information as the cut-off QCD Lagrangian
in terms of the original quark and gluon degrees of freedom. As in the
case of QCD, also the effective Lagrangian is amenable to numerical
studies from which the effective couplings can be extracted.

\section{Constituent Quarks and Mesonic Degrees of Freedom}

In this section we shall briefly outline the generalization of the
present effective Lagrangian technique to the case of the $SU(N_f)$
pseudoscalar multiplet. In contrast with the $U(1)$ case discussed
above, the flavour non-singlet axial currents are exactly conserved in
the limit of massless quarks. We need not be concerned with the
question of spontaneous chiral symmetry breaking here; the effects
will automatically be taken into account.

How do we introduce the appropriate collective fields for this
non-Abelian (flavoured) case? As before, the main input is the choice
of quantum numbers we wish to describe. We then start again with a
generating functional of QCD,
\bea
{\cal Z}[V,A] &=& \int\! {\cal D} [\bar{q},q] d\mu [G]\; e^{-\int\!
d^4x\; {\cal L}}
\cr {\cal L} &=& \bar{q} (\slash{\partial} -i\slash{G} -i\slash{V} -i\slash{A}
\gamma_5 )q + {\cal L}_{YM} ,
\eea
where we now consider external sources $V_\mu$ and $A_\mu$ that are
elements of $SU(N_f)$.

For technical reasons that will become clear shortly, it is most
convenient to introduce collective fields $\theta (x)$ by, $e.g.$,
purely left-handed transformations:
\beq
q_L(x) = e^{2i\theta (x)} \chi_L (x)\quad , \quad \bar{q}_L (x) =
\bar{\chi}_L (x) e^{-2i\theta (x)}
\label{eq:trL}
\eeq
$i.e.$ local phase transformations acting only on the left-handed
spinors:
\beq
q_L = P_+ q \quad , \quad \bar{q}_L = \bar{q} P_- \quad ,\quad P_\pm =
\frac{1}{2} (1\pm \gamma_5 )~.
\eeq
The transformation (\ref{eq:trL}) is a combination of an axial and a
pure vector gauge transformation. The latter can easily be corrected
for by a suitable opposite transformation; there are no complications
due to regularization, because we are using a scheme that preserves
vector gauge invariance.

Here $\theta (x)$ is understood to be an element of a Lie algebra, in
particular that of $SU(N_f)$,
\beq
\theta = \theta^a \lambda^a
\eeq
where $\lambda^a$ are the generators of $SU(N_f)$ with the convention
$tr\; \lambda^a \lambda^b = \delta^{ab}$. It is convenient to use the
combinations
\beq
L_\mu = V_\mu + A_\mu \quad , ~~~~~R_\mu = V_\mu - A_\mu
\eeq
instead of $V_\mu$ and $A_\mu$.

The transformation (\ref{eq:trL}) exhibits a change of the regularized
fermionic functional integral measure due to its handedness. In order
to calculate the corresponding contribution to the Lagrangian we again
use a Pauli-Villars regularization for reasons described in section
2. The generating functional now reads
\begin{eqnarray}
{\cal Z}_\Lambda [V,A] &=& \int\! {\cal D}_\Lambda [\bar{\chi},\chi ]
d\mu [G]\; e^{-\int\! d^4x\; {\cal L}'} \cr {\cal L}' &=& \bar{\chi}
\gamma_\mu (\partial_\mu -iG_\mu -iL_\mu^\theta P_+ -i R_\mu P_- )
\chi + {\cal L}_J + {\cal L}_{WZ} + {\cal L}_{YM}~,
\label{eq:ZnewL}
\end{eqnarray}
where only $L_\mu$ is modified as
\beq
L_\mu^\theta = U^\dagger L_\mu U + iU^\dagger \partial_\mu U~.
\eeq
We have now introduced the common symbol $U$ as
\beq
U(x) = e^{2i\theta (x)} \quad U^\dagger (x) = e^{-2i\theta (x)}~.
\eeq
Again, ${\cal L}_J$ and ${\cal L}_{WZ}$ are induced by the fermionic
measure.

The positive parity part can, as in the abelian case, be ordered as an
expansion in inverse powers of the ultraviolet cut-off $\Lambda$:
\beq
{\cal L}_J = \Lambda^2 {\cal L}_2 + {\cal L}_0 +
\frac{1}{\Lambda^2} {\cal L}_{-2} +
\frac{1}{\Lambda^4}{\cal{L}}_{-4} + \ldots
\eeq
where the first three terms are given by
\bea
{\cal L}_2 &=& \frac{N_c\kappa_2}{4\pi^2} tr_f A_\mu^{(s)} A_\mu^{(s)}
\mid_{s=1}^0 \cr {\cal L}_0 &=&
\frac{N_c}{8\pi^2} tr_f \biggl(-i F_{\mu\nu}^{(s)}
[A_\mu^{(s)} ,A_\nu^{(s)} ] + \frac{1}{3} D_\mu^{(s)} A_\nu^{(s)}
D_\mu^{(s)} A_\nu^{(s)} - \frac{2}{3} (A_\mu^{(s)} A_\mu^{(s)} )^2 \cr
&&\qquad + \frac{4}{3} A_\mu^{(s)} A_\nu^{(s)} A_\mu^{(s)} A_\nu^{(s)}
\biggr)
\mid_{s=1}^0\cr
{\cal L}_{-2} &=& \frac{\kappa_{-2}}{48\pi^2} tr_f
\biggl( N_c \partial^2 A_\mu^{(s)} \partial^2 A_\mu^{(s)} +
 A_\mu^{(s)} A_\mu^{(s)} tr_c G_{\nu\rho} G_{\nu\rho} + \ldots
\biggr)\mid_{s=1}^0~.
\label{eq:J+L}
\eea
Here, $tr_f$ and $tr_c$ denote the traces over flavour and colour
indices.  The terms omitted in ${\cal L}_{-2}$ and denoted by dots are
at least fourth order in $A_\mu^{(s)}$ and $V_\mu^{(s)}$.  The
expression for ${\cal{L}}_{-4}$ is quite lengthy, and we do not
reproduce it here.

The leading term of the negative parity part is the integrated
Bardeen-anomaly:
\bea
{\cal L}_{WZ} &=& \frac{i}{16\pi^2} \int_{1}^{0} ds
\;\epsilon_{\mu\nu\rho\sigma}\; tr_f tr_c \theta \biggl(
F_{\mu\nu}^{(s)} F_{\rho\sigma}^{(s)} + \frac{1}{3} A_{\mu\nu}^{(s)}
A_{\rho\sigma}^{(s)} \cr &&\qquad +
\frac{8i}{3} ( F_{\mu\nu}^{(s)} A_\rho^{(s)} A_\sigma^{(s)}
+ A_\mu^{(s)} F_{\nu\rho}^{(s)} A_\sigma^{(s)} + A_\mu^{(s)}
A_\nu^{(s)} F_{\rho\sigma}^{(s)} )
\cr && \qquad + \frac{32}{3} A_\mu^{(s)} A_\nu^{(s)} A_\rho^{(s)}
A_\sigma^{(s)} \biggr) + {\cal O} (\Lambda^{-2})~.
\label{eq:J-L}
\eea
The covariant derivatives and field strength tensors appearing in
eqs.  (\ref{eq:J+L}) and (\ref{eq:J-L}) are defined as
\begin{eqnarray}
{\cal D}_\mu A_\nu &=& \partial_\mu A_\nu -i [V_\mu ,A_\nu ]
\cr A_{\mu\nu} &=& {\cal D}_{[\mu} A_{\nu]} \cr F_{\mu\nu}
&=& \partial_{[\mu} V_{\nu]} - i[V_\mu , V_\nu ] - i[A_\mu ,A_\nu ]
\end{eqnarray}
and the transformed fields appearing in (\ref{eq:J+L}) and
(\ref{eq:J-L}) as
\bea
V_\mu^{(s)} &=& \frac{1}{2} \biggl( e^{-2is\theta} L_\mu e^{2is\theta}
+ i e^{-2is\theta} \partial_\mu e^{2is\theta} + R_\mu \biggr) \cr
A_\mu^{(s)} &=& \frac{1}{2} \biggl( e^{-2is\theta} L_\mu e^{2is\theta}
+ i e^{-2is\theta}
\partial_\mu e^{2is\theta} - R_\mu \biggr)~~.
\label{eq:AVs}
\eea
The parameter $s$ ranging from $0$ to $1$ thus defines a continuous
transformation which, for $s=1$, coincides with (\ref{eq:trL}).

If we now declare $\theta (x)$ a dynamical variable by integrating
over the invariant Haar measure $\int {\cal D}[U]$, a new local
non-Abelian gauge symmetry emerges:
\begin{eqnarray}
\chi_L (x) &\to& e^{2i\alpha (x)} \chi_L (x)\cr
\bar{\chi}_L (x) &\to& \bar{\chi}_L (x) e^{-2i\alpha (x)} \cr
U(x) &\to& U(x) e^{-2i\alpha (x)}
\label{eq:symL}
\end{eqnarray}
where $\alpha (x)$ is now a (local) transformation parameter in the
same representation of $SU(N_f)$ as $\theta (x)$.

As in the Abelian case, we discover an induced regularization,
actually due to the same operator $f^2$ as in (\ref{eq:f2}). This can
be seen by extracting from ${\cal L}_J$ the leading quadratic term in
$U^\dagger
\partial_\mu U$. In particular, it reads
\beq
{\cal L}_J = -\frac{1}{8} tr U^\dagger \partial_\mu U f^2 U^\dagger
\partial_\mu U +
\ldots
\eeq
The omitted terms are of higher order in $U^\dagger
\partial_\mu U$ or contain external sources.

As in the flavour-singlet case, the crucial step now is the choice of
gauge fixing. To derive a convenient form of the effective Lagrangian,
the optimal solution would be to find a suitable operator ${\cal O}_i$
in terms of the original quark fields $\bar{\psi}$ and $\psi$ such
that the quantum version of the same operator expressed in terms of
the chirally rotated quark fields $\bar{\chi},\chi$ and the correction
due to the chiral field $U$ separates. Then one can choose the very
convenient gauge in which ${\cal O}_i[\bar{\chi},\chi] = 0$. In the
flavour-singlet case, the chiral Jacobian precisely allows us to find
such a gauge; in that case there
is only a gauge fixing at the quantum level, and the above separation
of variables is guaranteed. Analogues of such a gauge fixing can be
found in the flavoured case too. Without
carrying such a gauge fixing through at this point, what will be the
rough form of the resulting gauge-fixed Lagrangian? It is interesting
to compare it with what has become known as the constituent chiral
quark model of Manohar and Georgi
\cite{MaGe} (see also the last part of ref.
\cite{Weinberg}).

In its original formulation \cite{MaGe} and with our conventions, the
chiral quark model is described in terms of a Lagrangian
\beq
{\cal{L}} = \bar{\psi}(i\slash{D} + \slash{v} + g_A\slash{a}\gamma_5 -
m)\psi + \frac{1}{4g^2} tr G_{\mu\nu}G_{\mu\nu} + \frac{1}{8} f^2 tr
\partial_{\mu}U^{\dagger}\partial_{\mu}U + \ldots
\label{eq:chqu}
\eeq
Here, $a_\mu$ and $v_\mu$ are given as
\beq
a_\mu = \frac{i}{2}\left( \xi^\dagger \partial_\mu \xi - \xi
\partial_\mu \xi^\dagger \right)~,~~~~
v_\mu = \frac{i}{2}\left( \xi^\dagger \partial_\mu \xi + \xi
\partial_\mu \xi^\dagger \right)
\eeq
and $U(x) = \xi(x)\xi(x)$. Up to an ordinary gauge transformation with
$\xi$ the expressions for $a_\mu , v_\mu$ coincide with those in
(\ref{eq:AVs}) for $s=1$. External sources have not been included. The
covariant derivative $D_{\mu}$ is just the usual $D_{\mu} =
\partial_{\mu} - iG_{\mu}$ of QCD. In the manner written above, this
appears to be the standard QCD Lagrangian, with the extra chiral-model
terms (and as yet undetermined chiral coupling $g_A$) added
artificially on top. The reason why this is argued not to overcount
the degrees of freedom \cite{MaGe} is that the quarks are to be viewed
as ``constituent" quarks in eq. (\ref{eq:chqu}). For example, the
quark mass term above is not the mass term of the current quarks, but
rather a number on the order of 350 MeV.

It would be tempting to identify our rotated and gauge-fixed fermion
fields $\bar{\chi}$ and $\chi$ with the constituent quarks of the
Lagrangian (\ref{eq:chqu}). Before doing this, we should, however,
recall that the fields $\bar{\chi},\chi$ have not been introduced for
the purpose of describing constituent quarks.\footnote{Collective
fields corresponding explicitly to constituent quarks can presumably
be introduced through suitable functional transformations as well. We
have not pursued this idea, though. To do so, we would require a more
precise definition of what is meant by constituent quarks in a
Lagrangian framework. See below.} Rather, they were a more or less
unavoidable step toward extracting the low-energy pseudoscalar degrees
of freedom from the QCD Lagrangian.

If we compare the suggested gauge-fixed effective Lagrangian discussed
above with the chiral quark model Lagrangian of eq.(\ref{eq:chqu}), we
note two additional differences: The presence in our formulation of a
Nakanishi-Lautrup field $b(x)$, and the ghosts $\bar{c} (x)$ and
$c(x)$. These fields are of course all artifacts of the gauge-fixing
formalism, and should be integrated out before a direct comparison can
be made. The Faddeev-Popov ghosts, although presumably non-trivially
coupled to the matrix $U(x)$ in the non-Abelian case, do not play a
very fundamental r\^{o}le. Integrating them out produces an infinite
series in derivatives of $U(x)$. These terms are of course crucial
for, $e.g.$, all axial Ward identities to be satisfied in the
rewritten theory, but we may think of them as only modifying the
expansion implied in eq. (\ref{eq:chqu}).

A much more crucial r\^{o}le is played by the field $b(x)$.
Integrating it out simply enforces the gauge-fixing constraint, and it
is precisely by means of this constraint that one has the possibility
of legitimately extracting the pion multiplet from the QCD
Lagrangian. Of course, by means of exact rewritings of a generating
functional for QCD, we do not learn about the preferred realization of
chiral symmetry. Without an at least partial solution of essential QCD
dynamics, we cannot see in what phase QCD may be realized in
Nature. But the rewrite may still make the physical picture more
transparent.

The point is the following. If we return to what would be our
effective Lagrangian, then, in the absence of explicit chiral symmetry
breaking in the form of current quark masses, we note that the pion
multiplet appears only through derivative couplings. This is one of
the standard features of chiral Lagrangians, implying the presence of
massless pseudoscalars -- the Goldstone bosons of the chiral symmetry
breaking $SU(N_f)_L
\times SU(N_f)_R \to SU(N_f)$. As we have stressed earlier, the
collective fields introduced by the technique described in this paper
are not tied to the possible existence of Goldstone modes. This is a
question of dynamics. We may or may not have spontaneous symmetry
breaking, and rewriting a generating functional of QCD adds nothing
new to this. But {\em if} we end up extracting strictly massless
pseudoscalars from the QCD Lagrangian, and if these fields are not
trivially decoupled, then the left-over QCD-like theory of chirally
rotated quark fields and gluons should be chirally
non-invariant.\footnote{We thank A. Manohar for emphasizing this
point.} This is basically the inverse of Goldstone's Theorem.

Certainly such a lack of chiral invariance in the truncated part of
our effective Lagrangian does not arise directly from an explicit
constituent mass of the chirally rotated quarks, the $\bar{\chi},\chi$
fields. If chiral symmetry is broken in the equivalent of the
truncated theory, the possibility exists that it occurs as an explicit
breaking due to the gauge-fixing constraint. This would be analogous
to the Abelian case in two dimensions. For $SU(2)$ the gauge-fixing
manifold would be a two-sphere. The gauge-fixing field $b(x)$ plays
the r\^ole of a momentum conjugate to the angles describing the
compact manifold. The global constraint on $b(x)$ is nothing else but
the quantization condition of the corresponding angular
momentum. Summation over the discrete values of the angular momentum,
as in the Abelian case, projects on a certain point of the compact
manifold and may thereby fix the values of the angles.  Such a
mechanism is possible in the $SU(N_f)$-case too, and could imply
chiral symmetry breaking in the truncated theory.

These considerations lead us to an interesting interpretation of the
term ``constituent quark". In our framework one starts with current
quarks. As one successively performs chiral rotations and the
associated non-trivial gauge-fixings, one effectively removes the
pseudoscalar degrees of freedom associated with the pairs of current
quarks-antiquarks. If chiral symmetry is broken, then at each chiral
rotation, the quarks become increasingly more ``constituent", and less
``current". This can be viewed as occurring through the fermion fields
moving through the background of the pseudo-Goldstone fields
$\theta(x)$. As there is never in this approach any double-counting of
degrees of freedom, these chirally rotated quark fields are, in the
phase of broken chiral symmetry, carrying less and less pseudoscalar
degrees of freedom. In two space-time dimensions, where the same
phenomenon should occur even in the abelian case, the chirally rotated
quarks can almost entirely disappear in this process: This is the result of
bosonization. It does not mean that constituent quarks cannot be given
a meaningful definition after bosonization, but they now appear in the
dual picture of a non-trivial soliton configuration
\cite{Ellis}. In four dimensions, there is no possibility of
constructing a topologically non-trivial configuration of the
colour-singlet pseudoscalar degrees of freedom which carries the
quantum numbers of a constituent quark. The chirally rotated quark
fields $\chi, \bar{\chi}$ are then the fields we wish to identify with
the constituent quarks. A related and perhaps complementary picture
has been suggested by Kaplan
\cite{Kaplan}, in which constituent quarks are viewed as Skyrmions in
``colour space''. Also such an approach can be pursued in a
collective-field framework similar to the one presented here for the
flavoured case. Bosonization of QCD$_2$ indeed confirms this picture
\cite{Ellis}.

\section{Conclusion}

We have presented a new technique for the derivation of effective
Lagrangians for long-distance dynamics, starting from an underlying
Lagrangian valid to much smaller distances. The derivation relies
heavily on a gauge-symmetric formulation that allows us to view
different field representations of a given theory as nothing but
different gauge slices.

As one application, this paper has focused on the classic problem of
deriving an effective low-energy Lagrangian for the flavour-singlet
sector of strong interactions, starting from QCD. The effective
Lagrangian we have presented is equivalent (in the sense of generating
functionals) to QCD with an explicit ultraviolet cut-off. It is
connected directly to QCD through a series of well-defined field
redefinitions in the cut-off theory, and in that sense is as
fundamental as the underlying cut-off QCD Lagrangian
itself. Technically, this is achieved by standard BRST quantization of
the equivalent gauge-symmetric formulation.

The scheme is obviously valid in larger generality. We have already
mentioned the case of two-dimensional physics, where it provides an
interpolation between purely fermionic and purely bosonic theories,
thereby extending the meaning of bosonization in two dimensions
\cite{us}. We have also indicated how it can be extended to
include the ``pion" multiplets of $SU(N_f)$ as extracted from QCD.  It
can straightforwardly be applied to any Lagrangian from which one
wishes to extract certain collective fields in an exact manner.
Particular cases of interest may involve, as in QCD, a focus on
Goldstone or pseudo-Goldstone fields described by coordinates on a
quotient space $G/H$, corresponding to a general spontaneous
symmetry-breaking pattern $G \to H$. Such a generalization may be of
interest in the study of, $e.g.$, technicolour theories. But as we
have repeatedly emphasized, collective fields may be of use much
beyond the description of nearly-massless pseudo-Goldstone bosons.

If we return to the effective $U(1)$ Lagrangian derived above, where
are the other pseudoscalars, the vector bosons, all the low-mass
resonances, the baryons, etc. in this formalism?  Since the effective
Lagrangians we have provided examples of are exact rewrites of cut-off
QCD, these other hadronic excitations below the cut-off are all
included.  We have not extracted them explicitly as collective fields,
and they are therefore simply contained in the dynamics of the
leftover (rotated) quark and gluon fields and their couplings to the
explicit meson sector -- just as also the lowest-mass pseudoscalars
are implicitly contained in the starting QCD Lagrangian. If it is
convenient to do so, one may introduce collective fields also for
these higher excitations. This, incidentally, does not exclude the
possibility that stable solitons with baryon quantum numbers can be
constructed out of the collective pion fields. To the extent that the
physical baryons indeed can be viewed as Skyrmions of meson
multiplets, it is quite possible that extracting pion degrees of
freedom from a generating functional of QCD also will entail a partial
extraction of baryonic excitations.  Following the rules laid out in
this paper, there will by construction never be any double counting of
the physical degrees of freedom.

\newpage
\bibliographystyle{unsrt}

\end{document}